\documentclass{aastex}
\usepackage[onecolumn,numberedappendix]{emulateapj5}

\usepackage{epsfig}

\shortauthors{T. Hosokawa \& K. Omukai}
\shorttitle{Evolution of Massive Protostars}

\begin{document}

\title{Evolution of Massive Protostars with High Accretion Rates
}
\author{Takashi Hosokawa and Kazuyuki Omukai}

\affil{Division of Theoretical Astronomy, 
National Astronomical Observatory of Japan, Mitaka, Tokyo 181-8588, Japan}
\email{hosokawa@th.nao.ac.jp; omukai@th.nao.ac.jp}

\begin{abstract}
Formation of massive stars by accretion requires a 
high accretion rate of $\dot{M}_\ast > 10^{-4}~M_{\odot}/{\rm yr}$
to overcome the radiation pressure barrier of the forming stars.
Here, we study evolution of protostars accreting at 
such high rates, by solving the structure of the central 
star and the inner accreting envelope simultaneously.
The protostellar evolution is followed starting from small initial cores 
until their arrival at the stage of the Zero-Age Main Sequence (ZAMS) stars. 
An emphasis is put on evolutionary features different 
from those with a low accretion rate of 
$\dot{M}_\ast \sim 10^{-5}~M_{\odot}/{\rm yr}$, 
which is presumed in the standard scenario for low-mass star formation. 
With the high accretion rate of 
$\dot{M}_\ast \sim 10^{-3}~M_{\odot}/{\rm yr}$, 
the protostellar radius becomes very large and exceeds 
$100~R_{\odot}$.
Unlike the cases of low accretion rates, 
deuterium burning hardly affects the evolution, and
the protostar remains radiative even after its ignition. 
It is not until the stellar
mass reaches $\simeq 40~M_{\odot}$ that hydrogen burning begins and
the protostar reaches the ZAMS phase, and this ZAMS arrival mass
increases with the accretion rate.
These features are similar to those of the first star formation 
in the universe, where high accretion rates are also expected, 
rather than to the present-day low-mass star formation. 
At a very high accretion rate of $> 3 \times 10^{-3}~M_{\odot}/{\rm yr}$,
the total luminosity of the protostar becomes so high 
that the resultant radiation pressure inhibits the 
growth of the protostars under steady accretion
before reaching the ZAMS stage. 
Therefore, the evolution under the critical accretion rate
$3 \times 10^{-3}~M_{\odot}/{\rm yr}$ gives the upper mass
limit of possible pre-main-sequence stars at $\simeq 60~M_{\odot}$.
The upper mass limit of MS stars is also set by the radiation
pressure onto the dusty envelope under the same 
accretion rate at $\simeq 250~M_{\odot}$.
We also propose that the central source enshrouded 
in the Orion KL/BN nebula has effective temperature and luminosity
consistent with our model, and is a possible candidate for such 
protostars growing under the high accretion rate. 
\end{abstract}


\section{Introduction}
\label{sec:intro}

Massive ($ > 8~M_{\odot}$) stars make significant impacts on 
the interstellar medium via various feedback processes:, e.g., 
UV radiation, stellar winds, and supernova explosions.
Such feedback processes sometimes trigger or regulate nearby 
star-formation activity.
Their thermal and kinetic effects are important factors
in the phase cycle of the interstellar medium.  
Strong coherent feedback by many massive stars
causes galactic-scale dynamical phenomena, such 
as galactic winds. Furthermore, massive stars
dominate the light from distant galaxies. 
The cosmic star formation rate is mostly estimated with
the light observed from massive stars \citep[e.g.,][]{Md96, HB06}. 
Despite such importance, the question of the formation process of 
massive stars still remains open.  
As for lower-mass stars, there is a widely-accepted formation scenario, 
where gravitational collapse of molecular cloud cores leads to subsequent 
mass accretion onto tiny protostars 
\citep[e.g.,][]{SAL87}.
If one directly applies this scenario to massive star
formation, however, some difficulties arise in the main 
accretion phase \citep[e.g., see][for recent reviews]{ZY07, MO07}.
The main difficulty in the formation of massive stars is the very
strong radiation pressure acting on a dusty envelope.
The radiative repulsive effect becomes quite strong at
the dust destruction front, where the accretion flow  
gets much of the outward momentum of radiation.  
Therefore, a necessary condition for massive star formation
is to overcome this barrier at the dust destruction front.
However, some theoretical work has shown that the accretion rate of 
$\dot{M}_\ast \sim 10^{-5}~M_{\odot}/{\rm yr}$ expected
of lower-mass protostars is too low to overcome the barrier
\citep[e.g.,][hereafter WC87]{WC87}.
Because of this difficulty, various scenarios different
from the standard accretion paradigm have been
proposed such as stellar mergers \citep[e.g.,][]{SPH00}
and competitive accretion \citep[e.g.,][]{BBZ98}.

An origin of this disputed situation is uncertainty concerning 
the initial condition for massive star formation. 
 This uncertainty partly originates from the scarcity of
massive stars; most of the massive star forming regions
are distant. 
In addition, the initial condition is easily disrupted
by feedback from newly formed massive stars.
Since the accretion rate reflects
the thermal state of the original molecular core, 
it is still uncertain which rates should be favored 
for growing high-mass protostars.
In fact, some observations of young high-mass sources
suggest high accretion rates.
Signatures of infall motion are detected with various 
line observations toward high-mass protostellar objects 
(HMPOs) and hyper-/ultra-compact H~II regions,
and the derived accretion rates are  
$10^{-4} - 10^{-3}~M_{\odot}/{\rm yr}$ 
\citep[e.g.,][]{Fl05, Bl06, KW06}.
Similar high accretion rates are also inferred by SED
fitting of hot cores \citep{OLD99} and HMPOs \citep{Fz07, KG08}.
Molecular outflows are also ubiquitous in high-mass 
star-forming regions, and estimated high mass outflow rates 
also suggest high accretion rates \citep[e.g.,][]{B02, Zh05}. 
 Such high accretion rates have the advantage of 
overcoming the radiation pressure barrier \citep{LS71, KH74}.
Theoretically, \citet{Nk00} have suggested a protostar growing 
at the very high accretion rate of 
$\sim 10^{-2}~M_{\odot}/{\rm yr}$ to explain the low radiation temperature
of scattered light from Orion KL region \citep{Mn98}.
\citet{MT02, MT03} have predicted that massive molecular
cloud cores, from which a few massive stars form, should be
dominated by supersonic turbulence, envisioning that 
such cores, if they exist, are embedded in a high-pressure environment.
\citet{KKM07} have simulated collapse of the turbulent cores
performing radiation-hydrodynamical calculations, and demonstrated
that the accretion rates attain more than $10^{-4}~M_{\odot}/{\rm yr}$
in their simulations.
Some recent observations are getting a close look at the initial
condition of massive star formation, and have found
strong candidates for high-mass pre-stellar
cores \citep[e.g.,][]{RSJ07, Mt07}. 
Further detailed observations will verify the scenario.

The main targets of this paper are protostars growing at the
high accretion rates of $\dot{M}_\ast \geq 10^{-4}~M_{\odot}/{\rm yr}$.
Detailed modeling of such protostars should be useful 
for future high-resolution observations and simulations of
massive star formation.
However, previous studies on such protostars are fairly limited.
The protostellar evolution has been well studied for low-mass 
($< 1~M_{\odot}$) and intermediate-mass ($< 8~M_{\odot}$) protostars
by detailed numerical calculations solving the stellar structure,
but these studies have focused on evolution at the low 
accretion rate of $\sim 10^{-5}~M_{\odot}/{\rm yr}$
(e.g., Stahler, Shu \& Taam 1980a,b, hereafter SST80a,b;
Palla \& Stahler 1990, 1991, hereafter PS91, 1992; 
Beech \& Mitalas 1994)
Maeder and coworkers have calculated the protostellar evolution
under accretion rates growing with the stellar mass, which
finally exceeds $10^{-4}~M_{\odot}/{\rm yr}$ 
\citep{BM96, BM01, NM00}, 
but it is still uncertain how the evolution changes with accretion 
rates, and which physical mechanisms cause differences.
Some authors have used polytropic one-zone models 
originally invented for low accretion rates
even at such a high rate \citep[e.g.,][]{Nk00, MT03, KT07}, but 
its validity remains obscure owing to the lack of detailed calculations.
By coincidence, the similar high accretion rates are expected
for protostars forming in the early universe \citep[e.g.,][]{ON98, Ys06}. 
This simply reflects high temperatures in primordial
pre-stellar clouds. Evolution of such primordial protostars has
been studied by \citet[][hereafter SPS86]{SPS86} and 
\citet{OP01, OP03}.  Therefore, our
other motivation is to investigate how the protostellar evolution 
changes with metallicities.

To summarize, our goal in this paper is to answer the following questions;
\begin{itemize}
\item What are the properties of massive protostars 
 (e.g., radius, luminosity, and effective temperature)
 growing at high accretion rates?
  Can we observe any signatures of their characteristic properties?
\item
 How different are high-mass protostars from low- and 
 intermediate-mass ones, for which lower accretion
 rates are presumed?
 How about the difference from the protostellar evolution of 
 primordial stars?
 Furthermore, if the protostellar evolution significantly varies with 
 accretion rates and metallicities, what causes such differences? 
\item 
  What are the consequences of protostellar evolution with
  high accretion rates for the formation and feedback processes
  of massive stars?
  How massive can a star get before its arrival at the 
  Zero-Age Main Sequence (ZAMS)?
  What is the maximum mass of stars that can be formed? 
\end{itemize}
In order to answer these questions, we solve the 
structure of accreting protostars with various accretion rates and 
metallicities with detailed numerical calculations
\citep[also see][for a similar effort]{YB08}.

The organization of this paper is as follows:
In \S~\ref{sec:num}, we briefly review the basic procedure to construct 
numerical models of accreting protostars and their surrounding envelopes.
The subsequent \S~\ref{sec:result} is the main part of this
paper, where our numerical results are presented.
First, we investigate the protostellar evolution of two fiducial 
cases with the accretion rates of $10^{-3}$ and 
$10^{-5}~M_{\odot}/{\rm yr}$ in \S~\ref{ssec:md_1em3} 
and \S~\ref{ssec:md_1em5}. After that, we show more general 
variations of protostellar evolution,
e.g., mass accretion rates in \S~\ref{ssec:md_dep},
and metallicities in \S~\ref{ssec:metal}.
In \S~\ref{ssec:edd}, we present the protostellar evolution
at very high accretion rates exceeding 
$10^{-3}~M_{\odot}/{\rm yr}$, and show that the steady accretion
is limited by a radiation pressure barrier acting on a gas
envelope.
We further discuss some implications based on our numerical results.
In \S~\ref{sec:stopacc}, we examine which feedback effects can
affect the growth of protostars for a given accretion rate.
We also explore some observational possibilities of detecting signatures 
of the supposed high accretion rates in \S~\ref{sec:obs}.
Finally, \S~\ref{sec:sum} is assigned to summary and conclusions.


\section{Numerical Modeling}
\label{sec:num}

\subsection{Outline}
\label{ssec:outline}

\begin{figure}[t]
  \begin{center}
\epsfig{ file=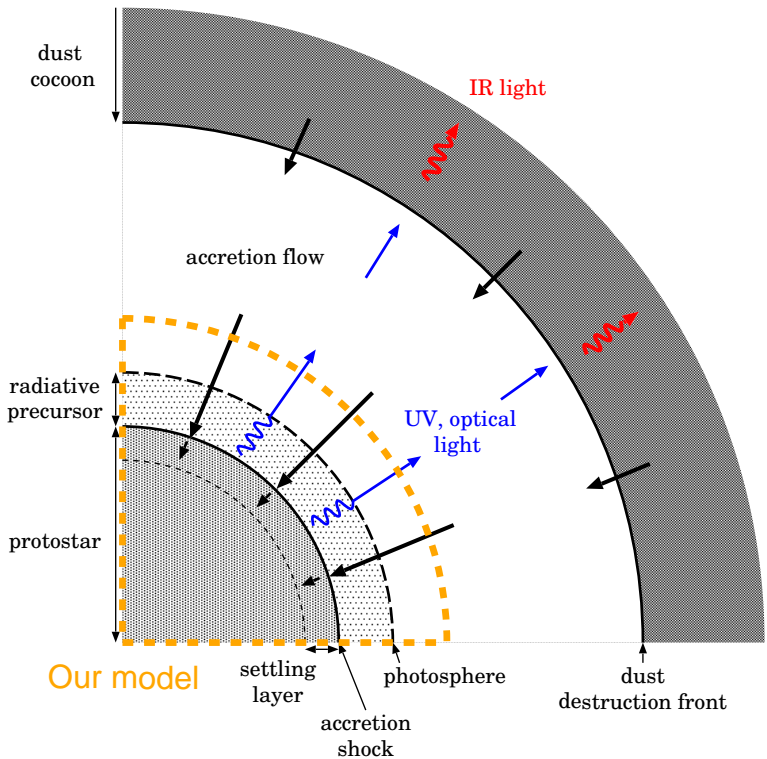,
         angle=0,
         width=4in}
\end{center}
\caption{
A schematic figure of a protostar and surrounding accretion 
flow.
The accretion shock forms at the stellar surface.
When the flow becomes optically thick before hitting on the shock, 
the photosphere forms outside the stellar surface.
The optically thick part of the accreting envelope is called 
the radiative precursor.
A dust cocoon surrounds the protostar at larger radius.
The inner boundary of the dust cocoon is a dust destruction front.
Most of the light from the protostar is absorbed at the 
dust destruction front once, and reemitted as infrared light.
In our calculations, we solve the detailed structure only of
the protostar and accretion flow far inside the dust destruction 
front, which are enclosed by the thick dashed line.
We discuss possible feedback effects on the outer dust cocoon
in \S~\ref{sec:stopacc}.
}
\label{fig:prost_schem}
\end{figure}


We employ the method of calculation developed by SPS86 and PS91. 
In this section, only the outline of modeling is overviewed.
A detailed description of the method is presented 
in Appendix~\ref{sec:method}.

As shown in Figure \ref{fig:prost_schem},
the whole system can be divided into two parts of different nature, 
i.e., an accreting envelope and a hydrostatic core.
This hydrostatic core is also called a protostar since it 
eventually grows into a star by accretion.
The outer part of the accreting envelope contains dust grains, 
while the warm ($\gtrsim 2000$K) part near the protostar
is dust-free as a result of their evaporation.
Despite the absence of the dust, the innermost dense part 
of the envelope becomes optically thick to gas continuum opacity 
and the photosphere appears outside the accretion shock front
in the case of a high accretion rate.
Such an inner envelope is called the radiative precursor. 
We here consider only the protostar and the radiative precursor
under a given constant accretion rate (see Fig.~\ref{fig:prost_schem}).
The dusty outer envelope is not included in our formulation,
although stellar feedback exerted on the envelope may be important in 
the last stage of accretion: the mass accretion can be terminated finally by 
radiation pressure or other protostellar feedback 
processes (e.g., stellar wind) exerted on the dusty envelope.
For the present, we presume that the protostar continues 
to grow at a given mass accretion rate and look for solutions 
of protostars with steady-state accretion. 
Discussion on feedback that halts the accretion is 
deferred to \S~\ref{sec:stopacc}.

We calculate protostellar evolution by constructing a
time sequence of quasi-steady structures of the protostar and 
accreting envelope. 
We solve the stellar structure equations for the protostar. 
For the accreting envelope, we adopt different treatments 
depending on whether or not it is opaque to the gas opacity. 
If the flow remains optically thin and no radiative precursor exists, 
the free-fall is assumed to be outside the star. 
For the opaque flow, on the other hand,
we solve the structure of the radiative precursor 
by using the equations for a steady-state flow.
At the outer boundary, which is taken to be at the photosphere, 
the flow is assumed to be in free fall.
After solving the protostar and accreting envelope individually, 
these solutions are connected at the accretion shock front 
by the radiative shock condition (e.g., SST80a).
The shooting method is adopted for solving radial structure.
This procedure is repeated until the required boundary
conditions are satisfied. 

We start calculation from a very small protostar, typically 
$M_{\ast, 0}= 0.01M_{\sun}$.
The initial models are constructed following SST80b 
(see Appendix~\ref{ssec:initial} for detail). 
Although this choice is rather arbitrary, the star converges
immediately to a certain structure appropriate for accretion 
at a given rate. 
Specific conditions of the initial models do not affect 
the evolution thereafter.   
The evolution is followed by increasing
the stellar mass owing to accretion step by step.
This procedure is repeated until the star reaches the ZAMS phase
after the onset of hydrogen burning.
In a few runs with very high accretion rates, however,
steady accretion becomes impossible before arrival at 
the ZAMS and the calculation is terminated at this moment.

\begin{table}[t]
\label{tb:md}
\begin{center}
Table 1. Calculated Runs and Input Parameters \\[3mm]
\begin{tabular}{l|ccccccc}
\hline
Run    & $\dot{M}_\ast~(M_{\odot}/{\rm yr})^a$ & 
         ${\rm [D/H]}~(10^{-5})^b$ &
         $Z^c$ & $M_{*,0}~(M_{\odot})^d$ & $R_{*,0}~(R_{\odot})^e$ & reference$^f$ \\
\hline
\hline
MD6x3     & $6 \times 10^{-3}$  & 2.5   & 0.02  & 0.5 & 43.9
                                & \S~\ref{ssec:edd}    \\
MD4x3     & $4 \times 10^{-3}$  & 2.5   & 0.02  & 0.3 & 33.0
                                & \S~\ref{ssec:edd}    \\
MD3x3     & $3 \times 10^{-3}$  & 2.5   & 0.02  & 0.2 & 27.8
                                & \S~\ref{ssec:edd}    \\
MD3x3-z0  & $3 \times 10^{-3}$  & 2.5   & 0.0   & 0.2 & 26.4
                                & \S~\ref{ssec:edd}    \\
MD3       & $10^{-3}$  & 2.5    & 0.02  & 0.05  & 15.5
                       &  \S~\ref{ssec:md_1em3}, \ref{ssec:md_dep}
                                               , \ref{ssec:metal}, \ref{ssec:edd}  \\
MD3-noD   & $10^{-3}$  & 0.0   & 0.02   & 0.05 & 15.5
                       &  \S~\ref{ssec:md_dep}   \\
MD3-z0    & $10^{-3}$  & 2.5   & 0.0   &  0.05 & 13.0
                       &  \S~\ref{ssec:metal}  \\
MD4       & $10^{-4}$  & 2.5   & 0.02  &  0.01 & 7.9
                       &  \S~\ref{ssec:md_dep}   \\
MD4-noD   & $10^{-4}$  & 0.0   & 0.02  &  0.01 & 7.9
                       &  \S~\ref{ssec:md_dep}   \\
MD5       & $10^{-5}$  & 2.5   & 0.02  &  0.01  & 3.7
                       &  \S~\ref{ssec:md_1em5}, \ref{ssec:md_dep} 
                        , \ref{ssec:metal} \\
MD5-noD   & $10^{-5}$  & 0.0   & 0.02  &  0.01  & 3.7
                       &  \S~\ref{ssec:md_1em5}, \ref{ssec:md_dep}   \\
MD5-dh1   & $10^{-5}$  & 1.0   & 0.02  &  0.01  & 3.7
                       &  Appendix \ref{ssec:dab}  \\
MD5-dh3   & $10^{-5}$  & 3.0   & 0.02  &  0.01  & 3.7
                       &  Appendix \ref{ssec:dab}  \\
MD5-z0    & $10^{-5}$  & 2.5   & 0.0   &  0.01  & 2.9
                       &  \S~\ref{ssec:metal}  \\
MD5-ps91$^g$  & $10^{-5}$  & 2.5   & 0.02  &  1.0 & 4.2
                       &  Appendix \ref{ssec:psini}  \\
MD6       & $10^{-6}$  & 2.5   & 0.02  &  0.01  &  1.6 
                       &  \S~\ref{ssec:md_dep}   \\
MD6-noD   & $10^{-6}$  & 0.0   & 0.02  &  0.01  & 1.6
                       &  \S~\ref{ssec:md_dep}   \\
\hline                           
\end{tabular}
\noindent
\end{center} 
$a$ : mass accretion rate, $b$ : initial number fractional abundance of
deuterium, $c$ : metallicity, $d$ : mass of initial core model,
$e$ : radius of initial core model,  
$f$ : subsections where numerical results of each run are presented, 
$g$ : initial model is the same as PS91

\end{table}

\subsection{Calculated Runs}

The calculated runs and their input parameters are listed in Table.1. 
Cases with a wide range of accretion rates are studied,
starting from $10^{-6}~M_{\odot}/{\rm yr}$ up to 
$6 \times 10^{-3}~M_{\odot}/{\rm yr}$. 
The lowest values of $10^{-6}- 10^{-5}~M_{\odot}/{\rm yr}$
are typical for low-mass star formation, 
while high rates $>10^{-4}~M_{\odot}/{\rm yr}$
are those envisaged in the accretion scenario of 
massive star formation. 
The initial deuterium abundance of [D/H] = 
$2.5 \times 10^{-5}$ is adopted as a fiducial value,  
following previous work (e.g., SST80a, PS91) for comparison. 
To assess the role of deuterium burning, 
we also calculate ``noD'' runs, 
where the deuterium is absent.
Most of the runs are for the solar metallicity $Z_{\sun}$(=0.02).
Two runs with $10^{-3}$ and $10^{-5} M_{\odot}/{\rm yr}$ are 
calculated for the metal-free gas (runs MD3-z0 and MD5-z0)
to see the effects of different metallicities.
For $10^{-5}~M_{\odot}/{\rm yr}$, variations in deuterium abundances 
(runs MD5-dh1 and MD5-dh3)
and in initial models (MD5-ps91) are studied and 
presented in Appendix \ref{ap:prev} for comparison with 
previous calculations.

Initial stellar mass in each run is taken to be sufficiently small, 
typically $M_{*,0} = 0.01~M_{\odot}$.
For high accretion rates $\dot{M}_\ast \geq 10^{-3}~M_{\odot}/{\rm yr}$,
somewhat more massive initial stars (see Table.1) are used 
since convergence of calculation was not achieved for lower mass ones.
The radii of initial models are also listed in Table.1.

\section{Protostellar Evolution with Different Accretion Rates}
\label{sec:result}

There are two important timescales for evolution of 
accreting protostars. The first one is the accretion timescale, 
\begin{equation}
t_{\rm acc} \equiv \frac{M_{\ast}}{\dot{M}_\ast} ,
\end{equation}
over which the protostar grows by mass accretion.
This is an evolutionary timescale of our calculations.
The second one is the Kelvin-Helmholtz (KH) timescale, 
\begin{equation}
t_{\rm KH} \equiv \frac{G M_{\ast}^2}{R_{\ast} L_{\ast}} ,
\end{equation}
over which the protostar loses energy by radiation.
The balance among these timescales is crucial to
the protostellar evolution.
When $t_{\rm KH} > t_{\rm acc}$, the radiative energy loss
hardly affects the protostellar evolution. 
The evolution is controlled only by the effect of mass accretion.
When $t_{\rm KH} < t_{\rm acc}$, on the other hand,
the evolution is mainly controlled by the radiative energy loss.

Luminosity from accreting protostars has two sources:
one is from the stellar interior, the other from the accretion
shock front.
In this paper, we call the former the interior luminosity $L_{\ast}$, 
and the latter the accretion luminosity,
\begin{equation}
L_{\rm acc} \equiv \frac{G \dot{M}_\ast M_{\ast}}{R_{\ast}} .
\label{eq:lacc}
\end{equation}
Total luminosity from the star is the sum of the interior 
and accretion luminosity: $L_{\rm tot} = L_{\ast} + L_{\rm acc}$.
Note that the balance between $t_{\rm KH}$ and $t_{\rm acc}$ also 
determines the dominant component of luminosity from accreting 
protostars, i.e., 
the ratio $L_{\rm acc}/L_{\ast}$ is equivalent to
$t_{\rm KH}/t_{\rm acc}$.

During the growth of the protostar, the KH timescale 
significantly decreases.  
As a result, the ratio $t_{\rm KH}/t_{\rm acc}$, which is
initially very large, falls below unity in the protostellar evolution. 
Below, we show that the decrease of $t_{\rm KH}/t_{\rm acc}$
indeed leads to a variety of evolutionary phases of accreting
protostars.

\subsection{Case with High Accretion Rate
$\dot{M}_\ast = 10^{-3}~M_{\odot}/{\rm yr}$}
\label{ssec:md_1em3}

\begin{figure}[t]
  \begin{center}
\epsfig{ file=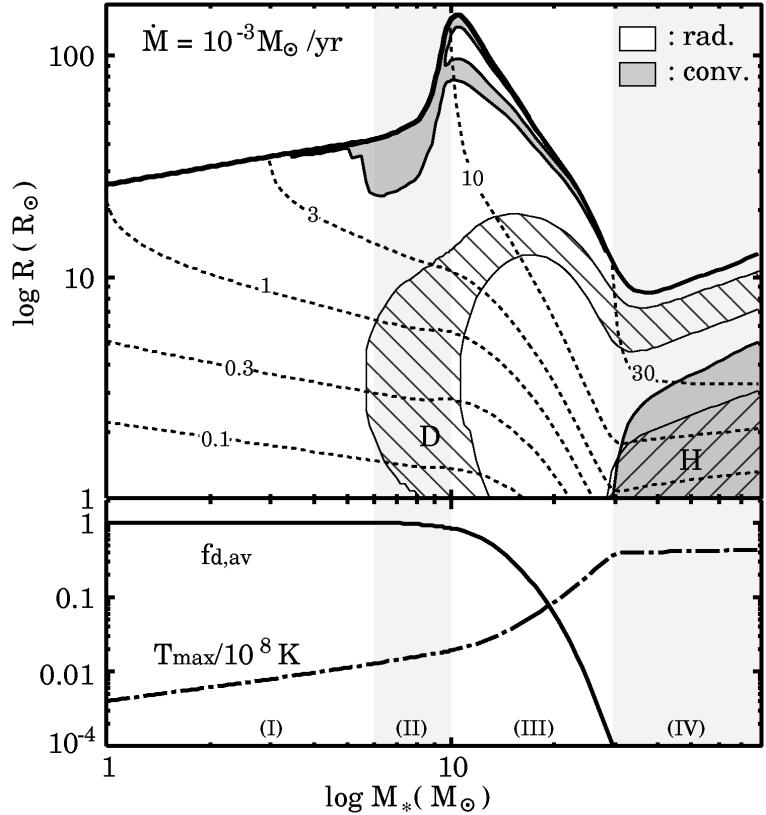,
         angle=0,
         width=4in}
\caption{ 
Evolution of a protostar with the accretion rate 
$\dot{M}_\ast = 10^{-3}~M_{\odot}/{\rm yr}$ (run MD3).
{\it Upper panel} : The interior structure of the protostar.
The thick solid curve represents the protostellar radius ($R_{\ast}$), 
which is the position of the accretion shock front. 
Convective layers are shown by gray-shaded area.
The hatched areas indicate layers of active nuclear burning,
where the energy production rate exceeds the steady rate
$L_{\rm D,st}/M_{\ast}$ for the deuterium burning, 
and $L_{\ast}/M_{\ast}$ for the hydrogen burning. 
The thin dotted curves represent the loci of 
mass coordinates ; $M = 0.1$, 0.3, 1, 3, 10, and 30~$M_{\odot}$.
{\it Lower panel} :
Evolution of the mass-averaged deuterium concentration,
$f_{\rm d, av}$ (solid line) and the maximum temperature within 
the star $T_{\rm max}$ (dot-dashed line).
In both panels, the shaded background shows the 
four evolutionary phases ;
(I) adiabatic accretion, (II) swelling, (III) Kelvin-Helmholtz
contraction, and (IV) main-sequence accretion phases.
}
\label{fig:str_fdtmax_1em3}
  \end{center}
\end{figure}

First, we see a protostar growing at a 
high accretion rate $\dot{M}_\ast = 10^{-3}~M_{\odot}/{\rm yr}$ (run MD3), 
which is relevant to massive star formation.
The upper panel of Figure \ref{fig:str_fdtmax_1em3} shows
evolution of the protostellar radius (thick solid) and 
the interior structure as a function of the protostellar mass. 
The mass increases monotonically with time $t$, 
$M_{\ast}=M_{\ast,0}+\dot{M}_\ast t$, and can be regarded as 
a time coordinate. 
The entire evolution can be divided into the following 
four phases according to their characteristic features;
(I) the adiabatic accretion (for $M_{\ast} \lesssim 6~M_{\odot}$), 
(II) swelling ($6~M_{\odot} \lesssim M_{\ast} \lesssim 10~M_{\odot}$),
(III) KH contraction 
($10~M_{\odot} \lesssim M_{\ast} \lesssim 30~M_{\odot}$), 
and (IV) main-sequence accretion ($M_{\ast} \gtrsim 30~M_{\odot}$) phases.
Below, we see the protostellar evolution in each phase.

\begin{figure}[t]
  \begin{center}
\epsfig{ file=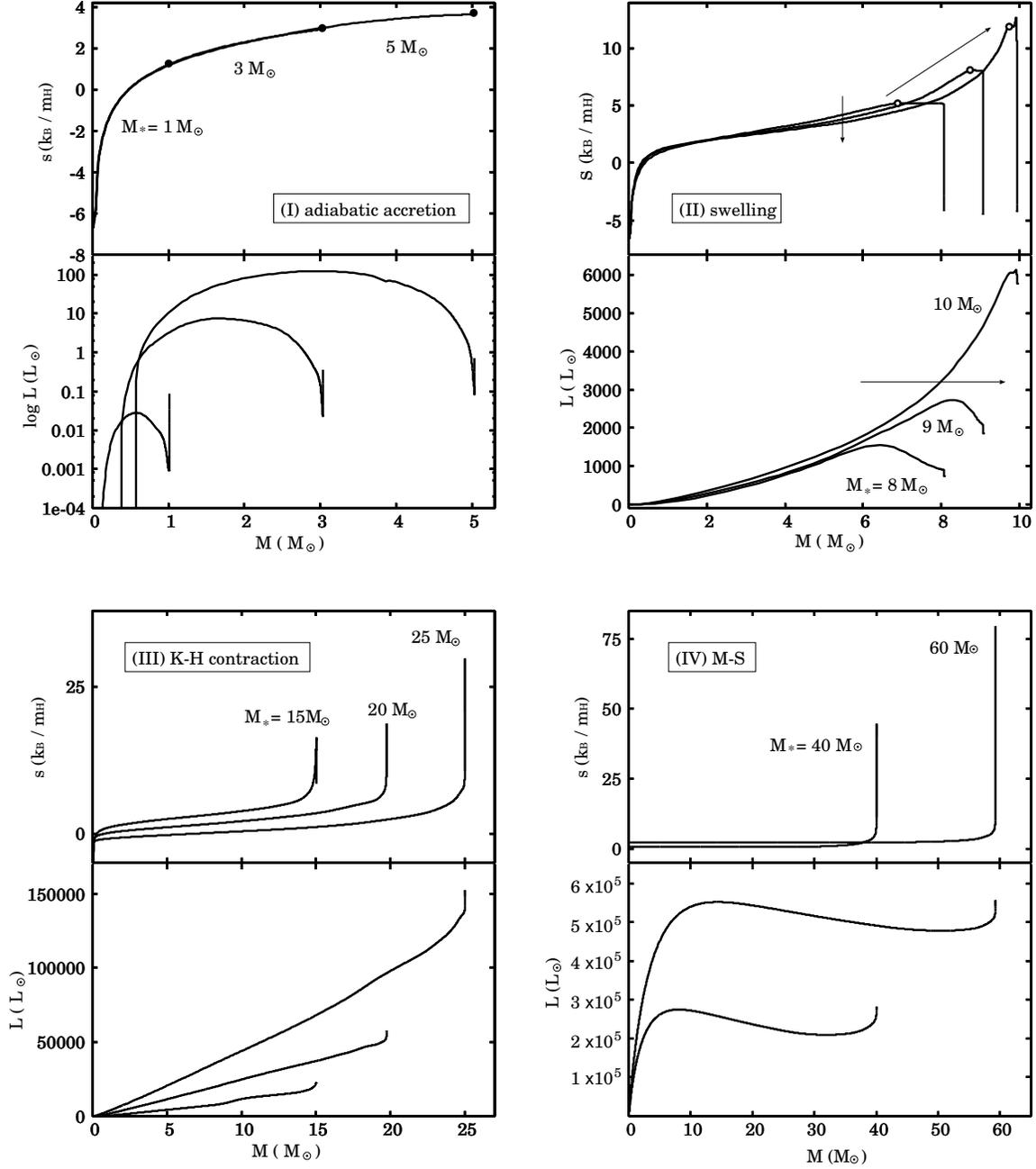,
         angle=0,
         width=6in}
\caption{Radial profiles of the specific entropy and
the luminosity at different epochs for the protostar with 
the accretion rate $\dot{M}_\ast = 10^{-3}~M_{\odot}/{\rm yr}$ (run MD3)
shown as functions of the mass coordinate $M$.
For each snapshot, total stellar mass ($M_{\ast}$) at its moment is labeled.
The four panels correspond to the four evolutionary phases, 
(I) adiabatic accretion (upper left), (II) swelling 
(upper right), (III) Kelvin-Helmholtz contraction (lower left), 
and (IV) main-sequence accretion (lower right) phases.
In the entropy profiles of the upper left panel, 
the filled circles denote the post-shock values.
In the upper right panel, 
the open circles indicate the bottom edges of convective layers.
}
\label{fig:slpf_1em3}
  \end{center}
\end{figure}
\begin{figure}[t]
\begin{center}
\epsfig{ file=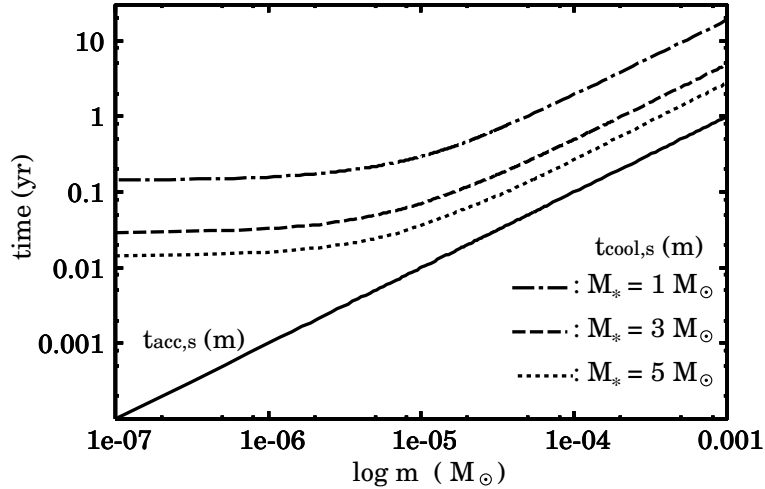,
         angle=0,
         width=4in}
\caption{
Comparison between the local accretion and cooling timescales
in the surface settling layer, defined by equations 
(\ref{eq:taccs}) and (\ref{eq:tcools}).
These timescales are shown as functions of 
the mass coordinate measured from the accretion shock front
$m \equiv M_{\ast} - M$.
The accretion timescale is presented by the solid line. 
The cooling timescales at $M_{\ast} = 1~M_{\odot}$, 
3~$M_{\odot}$, and 5~$M_{\odot}$ are presented
by the dot-dashed, dashed, and dotted lines, 
respectively. }
\label{fig:t_comp_1em3}
  \end{center}
\end{figure}
\begin{figure}[t]
  \begin{center}
\epsfig{ file=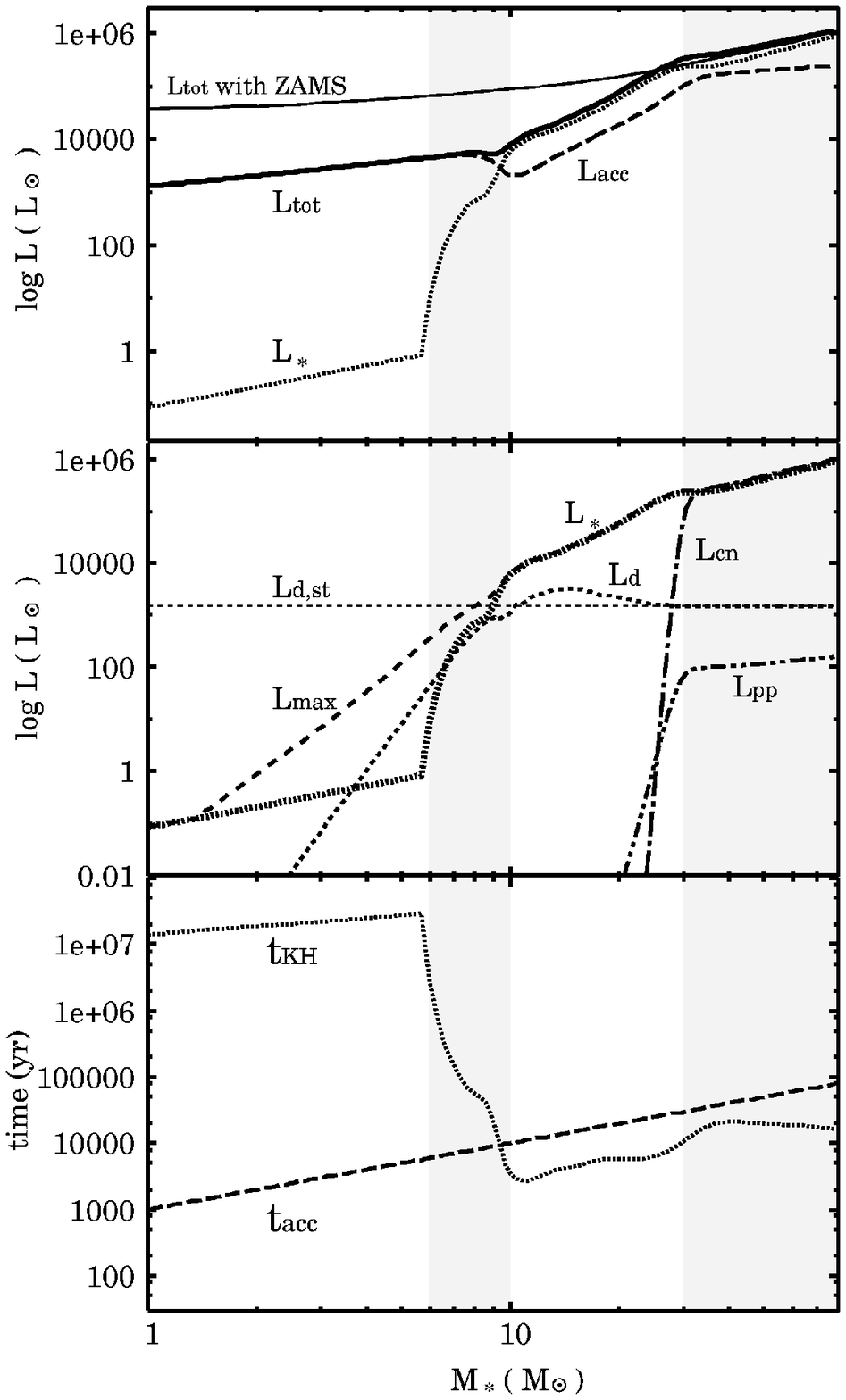,
         angle=0,
         width=4in}
\caption{ 
{\it Top panel} : Evolution of the interior luminosity
$L_{\rm *}$ (dotted), accretion luminosity 
$L_{\rm acc}$ (dashed), and total luminosity 
$L_{\rm tot} \equiv L_{\rm *} + L_{\rm acc}$
(solid) of an accreting protostar with 
$\dot{M}_\ast = 10^{-3}~M_{\odot}/{\rm yr}$ (run MD3).
The total luminosity calculated with radius and luminosity
of the ZAMS stars is also presented by the thin solid line.
{\it Middle panel} : Evolution of various contributions
to the interior luminosity of the protostar.
The thick dotted line represents the stellar luminosity,
$L_{\rm *}$ (also shown in the top panel).
The total burning rate of each nuclear reaction is
shown for the deuterium burning ($L_{\rm d}$, coarse dotted), 
pp-chain ($L_{\rm pp}$, dot-dot-dashed), and CN-cycle 
($L_{\rm CN}$, dot-dashed line). The horizontal coarse dotted line indicates 
the steady deuterium burning rate $L_{\rm d,st}$.
The maximum luminosity within the star is plotted with the coarse dashed
line.
{\it Bottom panel} : Evolution of the accretion timescale $t_{\rm acc}$
(dashed), and Kelvin-Helmholtz timescale $t_{\rm KH}$ 
(dotted) for the same protostar. 
In all panels, the shaded background shows the four evolutionary 
phases, as in Fig.~\ref{fig:str_fdtmax_1em3}. 
}
\label{fig:lum_enuc_tsc_1em3}
  \end{center}
\end{figure}

\paragraph{Adiabatic Accretion Phase}

Upper panels of Figure \ref{fig:slpf_1em3} present the evolution 
of entropy profile within the protostar. 
This shows that the entropy profiles at different epochs completely 
overlap: at a fixed mass element, the specific entropy 
remains constant. The entropy is originally generated at the accretion 
shock front, and embedded into the stellar interior.
Thus, during this earliest phase,  
the entropy profile inside the star just traces the history of 
entropy at the post-shock point. Owing to a high value of opacity, 
radiative heat transport is inefficient in the interior of the
protostar. This makes the interior luminosity small, then
$t_{\rm KH} \gg t_{\rm acc}$ in this phase 
(see Fig.~\ref{fig:lum_enuc_tsc_1em3} bottom panel, below).
Once the accreted matter settles into the opaque interior, the
specific entropy is just conserved.

The snapshots of the luminosity profile are also presented in 
lower panels of Figure \ref{fig:slpf_1em3}.
Near the surface is a spiky structure where the gradient
$\partial L/\partial M$ is very large owing to a sharp decrease in opacity there.
Although a significant entropy loss occurs in this luminosity spike as 
indicated by the relation 
$(\partial s/\partial t)_M \simeq - 1/T~(\partial L/\partial M)_t$ 
(from eq. (\ref{eq:ene})), 
this layer is extremely thin and the time for a mass element 
to pass though it is so short that the entropy hardly changes.
This can be confirmed by comparing the two timescales;
the local accretion time 
\begin{equation}
t_{\rm acc,s} (m) = \frac{m}{\dot{M}_\ast},
\label{eq:taccs}
\end{equation}
and local cooling time
\begin{equation}
t_{\rm cool,s} (m) = 
\frac{m~\delta s}{\int_0^m \frac1T \frac{\partial L}{\partial m} dm'},
\label{eq:tcools}
\end{equation}
where $m = M_{\ast} - M$ is the depth (in mass coordinate) of a mass shell 
from the accretion shock, 
and $\delta s$ denotes arbitrary small entropy change.
The former, $t_{\rm acc, s}$, is the time for 
a thin shell of mass $m$ at the surface to be replaced 
by the newly accreted material, while
the latter, $t_{\rm cool, s}$, is the time for
the same shell to lose the entropy $m~\delta s$ by outward
radiation. 
Comparison between these timescales in the settling layer is presented
in Figure \ref{fig:t_comp_1em3}.
where $\delta s = 0.1~k_{\rm B}/m_{\rm H}$ 
is adopted for numerical evaluation.
This indicates that $t_{\rm acc, s}$ is always shorter than $t_{\rm cool, s}$:
the accreted material is swiftly embedded in the interior before 
losing the entropy $\delta s$ by radiation.
Therefore, the adiabaticity remains valid 
in spite of the spike in luminosity profile.
Note that the short $t_{\rm acc,s}$ is a result of the high accretion
rate. 
We will see that the situation changes for much lower 
accretion rates in \S~\ref{ssec:md_1em5} and \ref{ssec:md_dep} below.

With increasing protostellar mass $M_{\ast}$, the accretion shock 
strengthens and the post-shock entropy increases.
Then, the interior entropy increases as well.
This causes a gradual expansion of the stellar radius in the adiabatic 
accretion phase (Fig.~\ref{fig:str_fdtmax_1em3}, upper panel).
Using the typical density and pressure within a star 
of mass $M_{\ast}$ and radius $R_{\ast}$ 
\citep[e.g.,][]{CG68};
\begin{equation}
\rho \sim \frac{M_{\ast}}{R_{\ast}^3}, 
\qquad P \sim G \frac{M_{\ast}^2}{R_{\ast}^4}.
\label{eq:typ_rhop}
\end{equation}
and the expression for specific entropy of ideal monatomic gas 
\begin{equation}
s = \frac{3 \cal{R}}{2 \mu} \ln \left( \frac{P}{\rho^{5/3}} \right) + const.
\label{eq:s_gene}
\end{equation}
where ${\cal R}$ is the gas constant and $\mu$ is the mean
molecular weight, 
the stellar radius and entropy is related as \citep{St88};
\begin{equation}
R_{\ast} \propto M_{\ast}^{-1/3} \exp\left[    
                            \frac{2 \mu}{3 {\cal R}} s
                           \right],
\label{eq:r_srel}
\end{equation} 
i.e., the stellar radius is larger for the higher entropy 
within the star. 
SPS86 have derived the mass-radius relation for 
protostars in the adiabatic accretion phase;
\begin{equation}
R_{\ast} \simeq  26~R_{\odot} \left( \frac{M_{\ast}}{M_{\odot}} \right)^{0.27} 
                     \left( \frac{\dot{M}_\ast}{10^{-3}~M_{\odot}/{\rm yr}} \right)^{0.41},
\label{eq:r_sps86}
\end{equation}
under the condition that opacity in the radiative precursor 
is dominated by H$^{-}$ bound-free absorption.
As this condition holds in our case,  
our calculated mass-radius relation is in a good agreement with
equation (\ref{eq:r_sps86}).
Equation (\ref{eq:r_sps86}) suggests the large radius
($> 10~R_\odot$) of the rapidly-accreting protostar.

The interior of the protostar remains radiative throughout this phase 
(Fig.~\ref{fig:str_fdtmax_1em3}, upper panel):
all the energy transport is via radiation.
This is in high contrast with low $\dot{M}_\ast$ 
($\sim 10^{-5}~M_{\odot}/{\rm yr}$) cases,
where most of the interior becomes convective owing to 
deuterium burning for $M_{\ast} \ga 0.4M_{\sun}$ 
(see Fig.~\ref{fig:str_fdtmax_1em5} below).
This can be attributed to a difference in the interior temperature.
From equation (\ref{eq:typ_rhop}), typical temperature within the star is  
\begin{equation}
T = \frac{\mu}{\cal{R}} \frac{P}{\rho} 
  \sim \frac{G}{\cal{R}} \frac{\mu M_{\ast}}{R_{\ast}}.
\label{eq:t_typ}
\end{equation}
Therefore, the large stellar radius leads to the
low temperature in the stellar interior.
In fact, the maximum temperature in this phase does not exceed 
the threshold for deuterium burning
(lower panel of Fig.~\ref{fig:str_fdtmax_1em3}).

Once the accreted matter has passed through the outermost layer
with the luminosity spike, the luminosity gradient becomes much milder.
The luminosity has a maximum at some radius:
outside the luminosity maximum, the gradient $\partial L / \partial M < 0$, 
while $\partial L / \partial M > 0$ inside.
This means that heat is removed from the deep interior
and absorbed in the outer layer.
This entropy transfer, however, remains small and does not
modify the entropy distribution significantly during this phase.
Efficiency of the outward entropy transfer is related to the
value of opacity.
In most of the stellar interior, major sources of opacity are
bound-bound and bound-free transitions of heavy elements, which 
approximately obey Kramers' law, 
$\kappa \propto \rho T^{-3.5}$ \citep{HHS62, Cl68}.
With the increase in stellar mass and then 
the interior temperature, the opacity decreases.
This results in steady increase of the outward heat flux 
with evolution.
In fact, as shown in the lower panels of Figure \ref{fig:slpf_1em3}, 
the amplitude of the luminosity increases with the growth of the 
protostar.
Also the maximum luminosity within the star $L_{\rm max}$ 
increases as a power-law function of $M_{\ast}$ 
as seen in the middle panel of Figure \ref{fig:lum_enuc_tsc_1em3}.
This dependence can be understood as follows:
for a star with radiative stratification and Kramers' opacity, 
the luminosity scales as
$L_{\rm rad} \propto M_{\ast}^{11/2} R_{\ast}^{-1/2}$ 
(e.g., Cox \& Giuli 1968).
We have confirmed that $L_{\rm max}$ roughly obeys,
\begin{equation}
L_{\rm max} \simeq 0.2~L_{\odot} 
                \left( \frac{M_{\ast}}{M_{\odot}}  \right)^{11/2}
                \left( \frac{R_{\ast}}{R_{\odot}}  \right)^{-1/2} 
\label{eq:l_max_a}
\end{equation}
in our calculations. 
Using equations (\ref{eq:r_sps86}) and (\ref{eq:l_max_a}), 
we obtain $L_{\rm max} \propto M_{\ast}^{5.4}$ for a constant
accretion rate.
When opacity becomes sufficiently small, the radiative heat transport 
becomes significant even in the deep interior.
Consequently, the entropy distribution and thus stellar structure are
drastically altered, which leads to the subsequent swelling phase.

\paragraph{Swelling Phase}

At the mass of about 6~$M_{\sun}$, 
the protostar begins to swell up dramatically.
The stellar radius suddenly increases by a factor of about three 
and eventually exceeds $100~R_{\odot}$. 
This swelling is caused by redistribution of entropy within the star.
As explained above, the outward heat flow continues to increase,
and ceases to be negligible.
Figure \ref{fig:slpf_1em3} shows that the entropy distribution changes 
significantly with time, i.e., with $M_{\ast}$: 
the entropy decreases in the deep interior, 
and increases in the outer layer. 
The extent of the inner entropy-losing region 
becomes larger and larger, and approaches the surface, 
as more of the interior becomes less opaque.

As seen in the bottom panels of Figure~\ref{fig:slpf_1em3}, 
the peak of the luminosity, which is the boundary between 
the entropy-losing interior and the outer absorbing layer, 
gradually moves toward the surface with increasing height.
This outward propagation of the luminosity peak is called the 
luminosity wave (SPS86). 
Only a thin outlying layer absorbs the 
entropy transported from the deep interior.
This rapid increase of the entropy near the surface 
causes swelling of the radius. 
Actually, only a small fraction of mass near the 
surface contributes to this swelling.
For example, at the maximum expansion of the radius 
$R_{\ast} \simeq 140~R_{\odot}$ at $M_{\ast} \simeq 10~M_{\odot}$, 
only 0.03\% of the total mass is contained 
in the outer layers of $R_{\ast} > 70~R_{\sun}$.

Note that the swelling is not due to the shell burning 
of deuterium.
\citet{PS90} have presented that a similar swelling occuring
in intermediate-mass protostars at the low accretion rate 
of $\dot{M}_\ast = 10^{-5}~M_{\odot}/{\rm yr}$, and concluded that
this is caused by the shell burning of deuterium. 
In our case, however, deuterium burning is not important 
for the swelling.
Even in the run without deuterium burning (MD3-noD),
the swelling is caused by redistribution of entropy in the same fashion
(also see \S~\ref{ssec:md_dep} below). 
During this phase, the deuterium burning occurs in the inner region
(Fig.~\ref{fig:str_fdtmax_1em3}, upper panel).
Despite the active deuterium burning, 
most of the stellar interior remains radiative.
Because of the higher entropy and thus lower density in the star, 
the opacity is lower in our case when the active deuterium burning
begins. 
This allows efficient radiative transport of the generated entropy 
\citep{St88}. 
The middle panel of Figure \ref{fig:lum_enuc_tsc_1em3} shows that
the maximum radiative luminosity $L_{\rm max}$ always exceeds
that by the deuterium burning $L_{\rm D}$.
This indicates a large capacity of radiative heat transport.
Also a thin convective layer near the surface 
(Fig.~\ref{fig:str_fdtmax_1em3}) is not brought about by the deuterium burning.
Although this layer receives the transported entropy from inside, 
the higher opacity due to ionization prevents the outward heat flow. 
This causes a negative entropy gradient and then convection near the surface.

\paragraph{Kelvin-Helmholtz Contraction}

With the approach of the luminosity wave to the surface, 
an increasing amount of energy flux escapes from the star 
without being absorbed inside.
This results in a significant increase in the interior luminosity 
$L_{\ast}$ for $M_{\ast} > 5 M_{\odot}$ (Fig.~\ref{fig:lum_enuc_tsc_1em3}) 
and thus the shorter KH timescale $t_{\rm KH}$ ($\propto 1/L_{\ast}$) 
of the star.
The arrival of the luminosity wave at the stellar surface marks
the moment that all parts of the star begin to lose heat.
Almost simultaneously, $t_{\rm KH}$ becomes as short as the accretion
timescale $t_{\rm acc}$ (Fig.~\ref{fig:lum_enuc_tsc_1em3}). 
While maintaining the virial equilibrium against the radiative energy loss,
the protostar turns to contraction.
This is the so-called KH contraction phase. 
The epoch of the radial turn-around is roughly estimated 
by equating $t_{\rm acc}$ and 
$t_{\rm KH,lmax} \equiv G M_{\ast}^2 / R_{\ast} L_{\rm max}$, 
where $L_{\rm max}$
is given by equation (\ref{eq:l_max_a}),
\begin{equation}
M_{\rm *,rmax} = 10.8~M_{\odot} \left(  
                           \frac{\dot{M}_\ast}{10^{-3}~M_{\odot}/{\rm yr}}
                          \right)^{2/9}.
\label{eq:m_rmax}
\end{equation}
In the above, we have omitted a weak dependence on $R_{\ast}$ and just 
substituted $R_{\ast} \sim 10~R_{\odot}$ for simplicity. 
During the contraction of the protostar, the KH timescale remains slightly
shorter than the accretion timescale. 
Since the ratio $t_{\rm KH}/t_{\rm acc}$, or 
equivalently $L_{\ast}/L_{\rm max}$,
is less than unity, the interior luminosity exceeds the accretion 
luminosity and becomes the dominant source of the total luminosity 
of the protostar (upper panel of Fig.~\ref{fig:lum_enuc_tsc_1em3}).

Since the stellar interior remains radiative in spite of the 
deuterium burning, the accreted deuterium is not transported
to the deep interior by convection.
Deuterium in the deep interior is thus soon consumed. 
Thereafter the deuterium burning continues in a shell-like outer layer
(Fig.~\ref{fig:str_fdtmax_1em3}, upper panel).
This surface burning proceeds at a rate slightly exceeding
the steady-burning one \citep[e.g.,][]{PS90}, 
\begin{equation}
L_{\rm D, st} \equiv \dot{M}_\ast \delta_{\rm D} = 
                1500~L_{\odot} \left( 
                               \frac{\dot{M}_\ast}{10^{-3}~M_{\odot}/{\rm yr}} 
                             \right)
                             \left(
                               \frac{[ {\rm D/H} ]}{2.5 \times 10^{-5}} 
                             \right), 
\label{eq:l_dst}
\end{equation}
where $\delta_{\rm D}$ is the energy available from
the deuterium burning per unit gas mass.
Consequently, the mass-averaged deuterium concentration $f_{\rm d,av}$ 
significantly decreases during the KH contraction
(Fig.~\ref{fig:str_fdtmax_1em3}, lower panel).

\paragraph{Arrival at a Main-sequence Star Phase}

With the KH contraction, the interior temperature increases gradually.
At $M_{\ast} \simeq 20~M_{\odot}$ the maximum temperature reaches $10^7$K
and the nuclear fusion of hydrogen begins (Fig.~\ref{fig:str_fdtmax_1em3}).
Although the hydrogen burning is initially dominated by the 
pp-chain reactions, energy generation by the CN-cycle reactions 
immediately overcome this. 
The energy production by the CN-cycle compensates 
radiative loss from the surface at $M_{\ast} \simeq 30~M_{\odot}$ 
(Fig.~\ref{fig:lum_enuc_tsc_1em3}, middle), where the 
KH contraction terminates.
A convective core emerges owing to the rapid entropy generation 
near the center.
The stellar radius increase after that, 
obeying the mass-radius relation of main-sequence stars.

\begin{figure}[t]
\begin{center}
\epsfig{ file=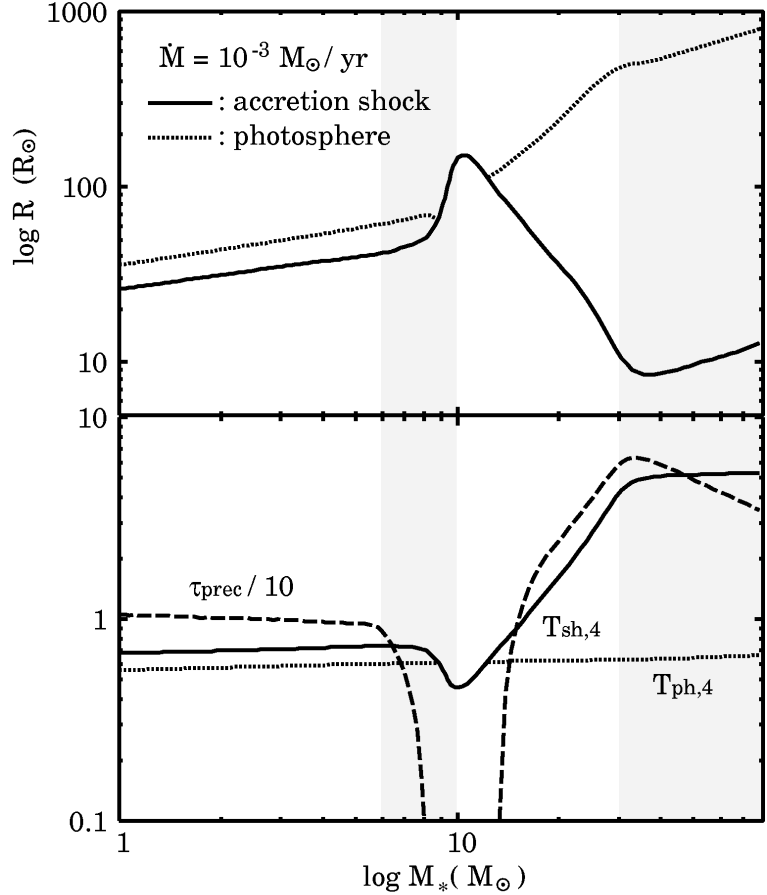,
         angle=0,
         width=4in}
\caption{ 
{\it Upper panel} :
Positions of the accretion shock front (solid line), 
and photosphere (dotted line) of a growing protostar 
with $\dot{M}_\ast = 10^{-3}~M_\odot/{\rm yr}$
(run MD3).
The layer between the accretion shock front and photosphere
corresponds to the radiative precursor.
{\it Lower panel} : 
the optical depth within the radiative precursor $\tau_{\rm rec}$, 
and effective temperatures at the photosphere $T_{\rm ph}$
and accretion shock front 
$T_{\rm sh} \equiv (L_{\ast}/4 \pi R_{\ast}^2 \sigma)^{1/4}$. 
The temperatures are normalized as $T_4 = (T/10^4 {\rm K})$.
The shaded background shows the four evolutionary phases,
as in Fig.~\ref{fig:str_fdtmax_1em3}. 
}
\label{fig:prec_1em3}
\end{center}
\end{figure}

\paragraph{Evolution of the Radiative Precursor}

Figure \ref{fig:prec_1em3} shows the evolution of 
protostellar ($R_{\ast}$) and photospheric ($R_{\rm ph}$) radii.
Except in a short duration, the photosphere is formed outside 
the stellar surface: the radiative precursor persists 
throughout most of the evolution.
In the adiabatic accretion phase, the photospheric radius is
slightly outside the protostellar radius and increases gradually with 
the relation $R_{\rm ph} \simeq 1.4 R_{\ast}$, which is derived 
analytically by SPS86.
Although the precursor temporarily disappears in the swelling phase
around $M_{\ast} \simeq 10 M_{\odot}$, 
it emerges again in the subsequent KH contraction phase.
In the KH contraction phase, the precursor extends spatially and
becomes more optically thick.
Due to the extreme temperature sensitivity of H$^-$ bound-free 
absorption opacity, the region in the accreting envelope with 
temperature $\ga 6000$K becomes optically thick.
Thus, the photospheric temperature $T_{\rm ph}$ remains at $6000-7000$K 
throughout the evolution (Fig.~\ref{fig:prec_1em3}, lower panel).
Since the photospheric radius is related to the total luminosity by 
\begin{equation}
R_{\rm ph}=\left(L_{\rm tot}/4\pi \sigma T_{\rm ph}^4 \right)^{1/2},
\end{equation}
for the constant $T_{\rm ph}$,
the rapid increase in the luminosity in the contraction phase 
causes the expansion of the photosphere.

\subsection{Case with Low Accretion Rate
            $\dot{M}_\ast = 10^{-5}~M_{\odot}/{\rm yr}$}
\label{ssec:md_1em5}

Next, we revisit the evolution of an accreting protostar
under the lower accretion rate of $\dot{M}_\ast = 10^{-5}~M_{\odot}/{\rm yr}$
(run MD5). 
Although this case has been extensively studied by previous 
authors, a brief presentation should be helpful
to underline the effects of different accretion rates on 
protostellar evolution. 
Different properties of the protostar 
even at the same accretion rate owing to updates 
from the previous works,   
e.g, initial models and opacity tables, 
will also be described.
For a thorough comparison between our and some previous calculations,
see appendix~\ref{ap:prev}.

\begin{figure}[t]
\begin{center}
\epsfig{ file=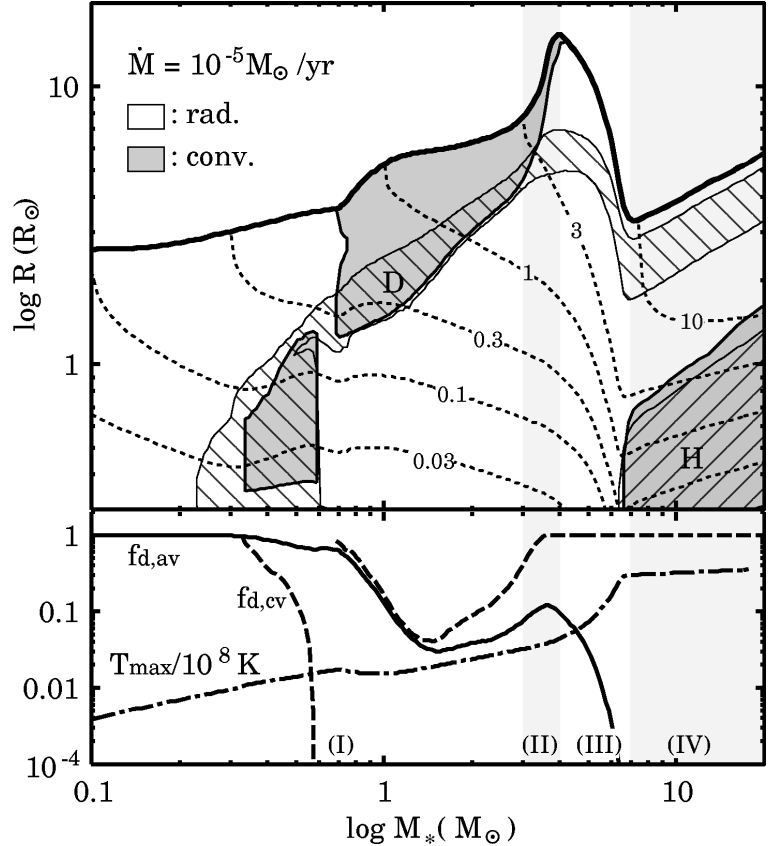,
         angle=0,
         width=4in}
\caption{ Same as Fig.~\ref{fig:str_fdtmax_1em3} but for the lower
accretion rate of $\dot{M}_\ast = 10^{-5}~M_{\odot}/{\rm yr}$ (run MD5). 
In the lower panel, deuterium concentration in the convective layer
$f_{\rm d,cv}$ is also presented.
In both upper and lower panels, the shaded background shows the 
four evolutionary phases ;
(I) convection, (II) swelling, (III) Kelvin-Helmholtz
contraction, and (IV) main sequence accretion phases}
\label{fig:str_fdtmax_1em5}
\end{center}
\end{figure}

The lower accretion rate means the longer accretion timescale
$t_{\rm acc}\propto 1/\dot{M}_\ast$:
even at the same protostellar mass $M_{\ast}$, the evolutionary 
timescale is longer in the case of low $\dot{M}_\ast$.
Therefore, the protostar has ample time to lose heat before 
gaining more mass.
This results in the lower entropy and thus a smaller radius
at the same protostellar mass in the low $\dot{M}_\ast$ case 
as shown in the upper panel of Figure \ref{fig:str_fdtmax_1em5}.
Although with the smaller value, the overall evolutionary 
features of the high and low $\dot{M}_\ast$ radii are
similar.

Again, the protostellar evolution can be divided 
into four characteristic stages, i.e., 
(I) convection ($M_{\ast} \lesssim 3~M_{\odot}$), 
(II) swelling ($3~M_{\odot} \lesssim M_{\ast} \lesssim 4~M_{\odot}$),
(III) KH contraction ($4~M_{\odot} \lesssim M_{\ast} \lesssim 7~M_{\odot}$),
and (IV) main-sequence accretion ($M_{\ast} \gtrsim 7~M_{\odot}$) phases.
Note that the first phase is now the convection phase instead of 
the adiabatic accretion. 
In the low $\dot{M}_\ast$ case, the deuterium burning 
drastically affects the protostellar evolution in early phases 
\citep[e.g.,][]{St88}.
Since the evolution in the phases (III) and (IV) is similar to 
the previous high-$\dot{M}_\ast$ case (\S~\ref{ssec:md_1em3}), 
we emphasize the two early phases (I) and (II) in the following. 

\begin{figure}[t]
  \begin{center}
\epsfig{ file=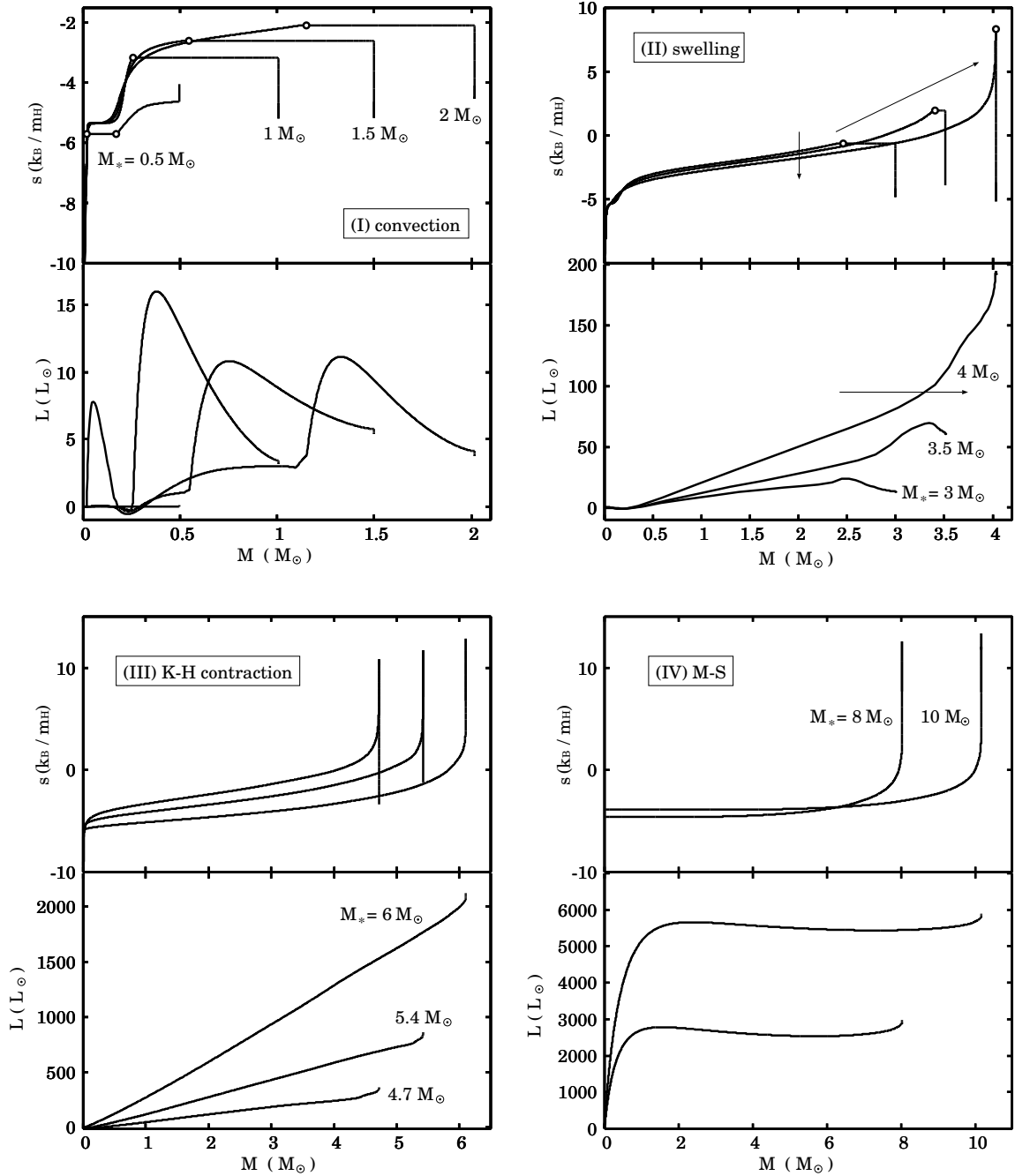,
         angle=0,
         width=6in}
\caption{Same as Fig.~\ref{fig:slpf_1em5} but for the lower accretion
rate of $\dot{M}_\ast = 10^{-5}~M_{\odot}/{\rm yr}$ (run MD5).
 }
\label{fig:slpf_1em5}
  \end{center}
\end{figure}
\begin{figure}[t]
\begin{center}
\epsfig{ file=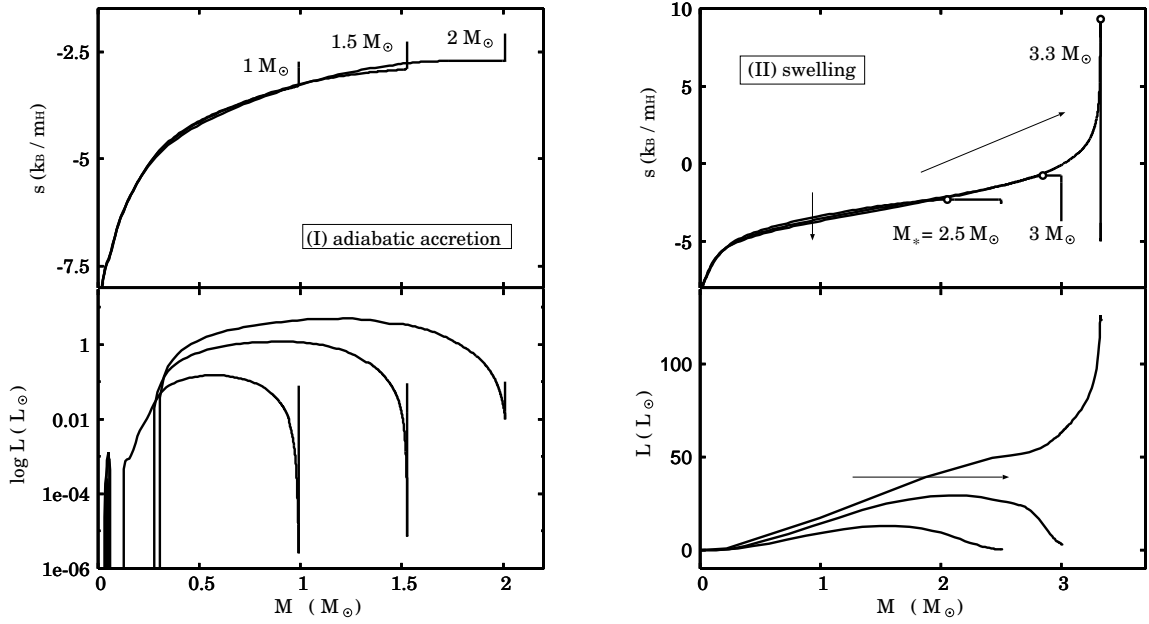,
         angle=0,
         width=6in}
\caption{Same as Fig.~\ref{fig:slpf_1em5} but for switching off
deuterium burning in the calculation (run MD5-noD). 
Only profiles in the early adiabatic accretion phase 
and swelling phase are presented.
}
\label{fig:slpf_1em5_nod}
\end{center}
\end{figure}
\begin{figure}
  \begin{center}
\epsfig{ file=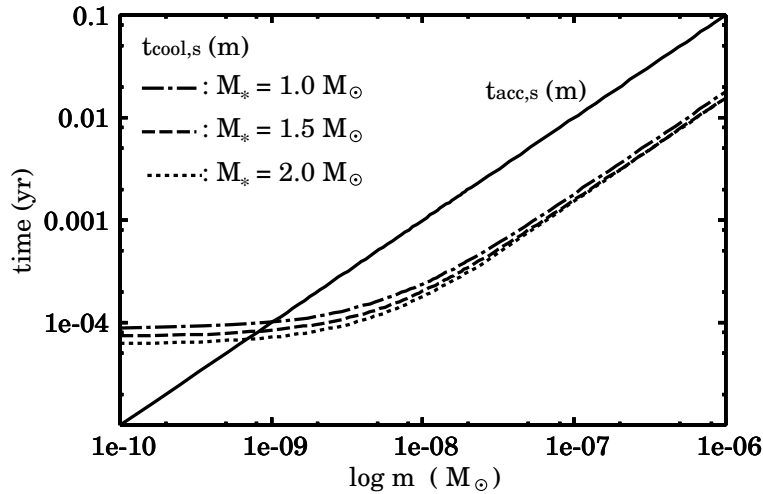,
         angle=0,
         width=4in}
\caption{Same as Fig.~\ref{fig:t_comp_1em3} but for the case with
the lower accretion rate of $\dot{M}_\ast = 10^{-5}~M_{\odot}/{\rm yr}$ 
and without deuterium burning (run MD5-noD). 
The local cooling timescales given by equation (\ref{eq:tcools}) 
at $M_{\ast} = 1.0$ (dot-dashed), 
1.5 (dashed), and 2.0~$M_{\odot}$ (dotted line) are presented.
}
\label{fig:t_comp_1em5}
  \end{center}
\end{figure}
\begin{figure}[t]
\begin{center}
\epsfig{ file=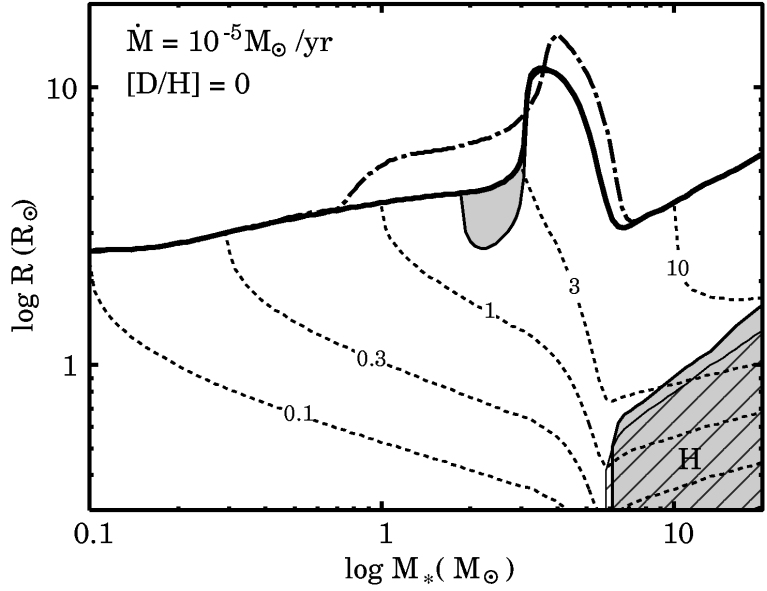,
         angle=0,
         width=4in}
\caption{ Same as the upper panel of Fig.~\ref{fig:str_fdtmax_1em5}
but for the case without the deuterium burning
([D/H] = 0, run MD5-noD).
The dot-dashed line represents the mass-radius relation 
for the case with the fiducial deuterium abundance, 
[D/H] = $2.5 \times 10^{-5}$ 
(run MD5, also see Fig.~\ref{fig:str_fdtmax_1em5}).
}
\label{fig:str_1em5_nod}
\end{center}
\end{figure}
\begin{figure}[t]
  \begin{center}
\epsfig{ file=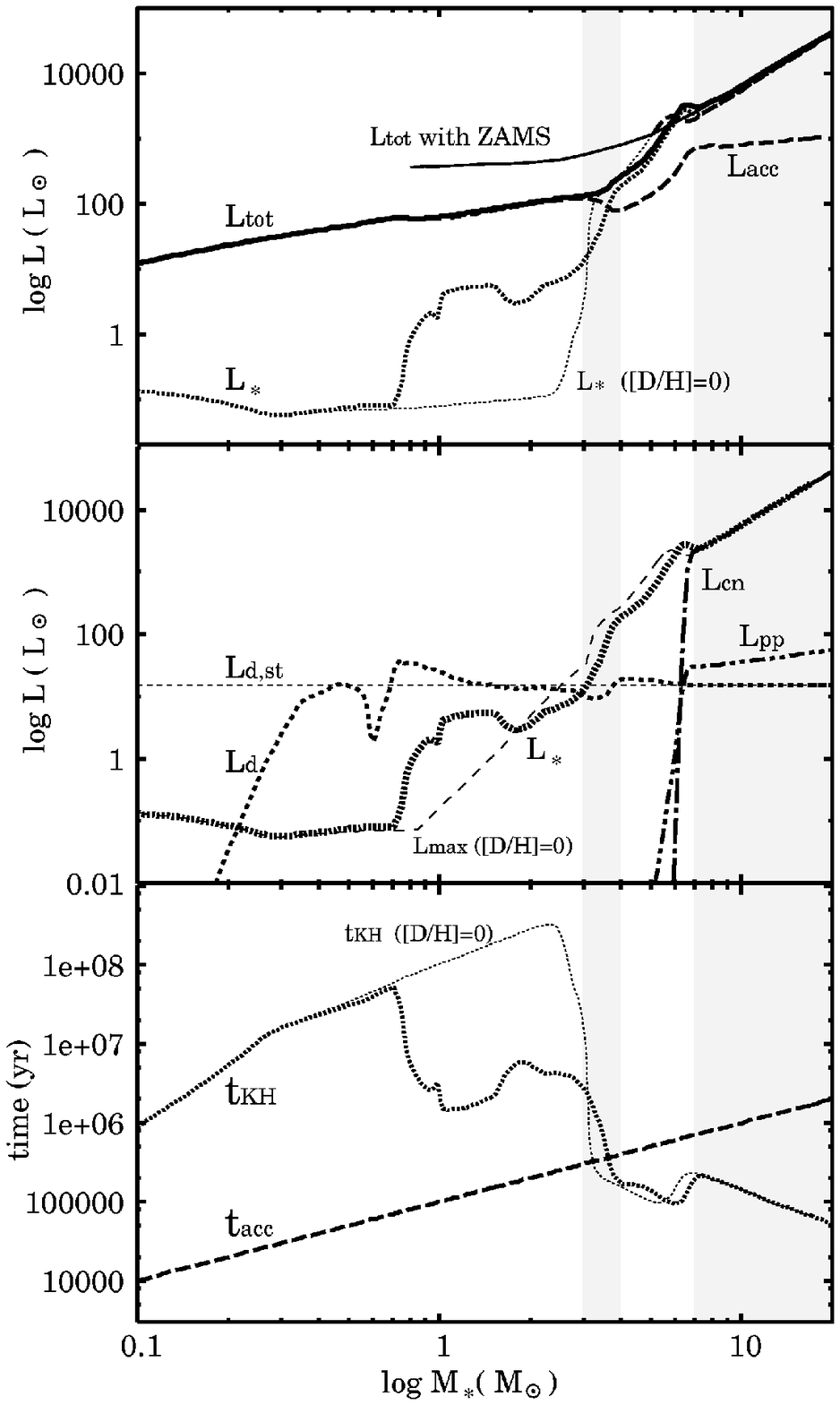,
         angle=0,
         width=4in}
\caption{ 
Same as Fig.~\ref{fig:lum_enuc_tsc_1em3} but for 
the lower accretion rate of 
$\dot{M}_\ast = 10^{-5}~M_{\odot}/{\rm yr}$ (run MD5). 
In each panel, the stellar luminosity 
$L_{\rm *}$ ({\it Top panel}, thin dotted line), 
maximum luminosity within the star $L_{\rm max}$ 
({\it Middle panel}, thin coarse dashed line),
and Kelvin-Helmholtz time $t_{\rm KH}$ 
({\it Bottom panel}, thin dotted line) for the case without 
deuterium burning (run MD5-noD) are presented for comparison.
In all panels, the shaded background shows the four characteristic 
phases, as in Fig.~\ref{fig:str_fdtmax_1em5}. 
}
\label{fig:lum_enuc_tsc_1em5}
  \end{center}
\end{figure}

\paragraph{Convection Phase}

In the low $\dot{M}_\ast$ case, the D burning begins in an earlier phase 
and has a more profound effect on the evolution
than in the high-$\dot{M}_\ast$ case (run MD3).
As a result of the smaller radius, the temperature, as well as density, 
is higher in the low $\dot{M}_\ast$ case 
(cf. lower panels of Fig.~\ref{fig:str_fdtmax_1em3} and 
Fig.~\ref{fig:str_fdtmax_1em5}).
The maximum temperature reaches $10^6$~K as early as 
$M_{\ast} \simeq 0.2~M_{\odot}$ 
and active deuterium burning begins subsequently
(Fig.~\ref{fig:str_fdtmax_1em5}), while in the high-$\dot{M}_\ast$ case 
the D ignition does not occur until $M_{\ast} \simeq 6~M_{\odot}$ 
(Fig.~\ref{fig:str_fdtmax_1em3}).
Unlike in the high-$\dot{M}_\ast$ case,
the deuterium burning makes most of the interior convective.
This is a result of high density and thus high opacity 
at the ignition of deuterium burning, occurring 
at a fixed temperature of $\simeq 10^{6}$ K.
Since high opacity hinders the radiative heat transport,
entropy generated by deuterium burning cannot be transported radiatively.
This necessarily gives rise to convection.

Figure \ref{fig:str_fdtmax_1em5} shows that
the extent of the convective layers experiences a complicated history.
At $M_{\ast} \simeq 0.3~M_{\odot}$, a convective layer appears 
for the first time slightly off-center. 
Inside the convective layer, entropy is homogenized and its value 
increases owing to the D burning (Fig.~\ref{fig:slpf_1em5}, I).
With this increased entropy, the outer layer is gradually incorporated 
into the convective region. 
Fresh deuterium is supplied from the newly incorporated layer  
and spreads over the convective region, and finally is consumed 
by nuclear burning at the bottom of the region. 
However, the expansion of the convective region cannot be maintained 
if the supply of deuterium becomes insufficient.
Without the fuel, the deuterium concentration in the convective region 
$f_{\rm d,cv}$ drops and the nuclear burning ceases: 
the convective region disappears.   
This occurs at $M_{\ast} \simeq 0.6~M_{\odot}$ 
(Fig.~\ref{fig:str_fdtmax_1em5}, upper).
At $M_{\ast} \simeq 0.7~M_{\odot}$, another episode of deuterium burning 
begins just outside the first convective region, where fresh deuterium 
still remains.
The second convective region develops there and immediately reaches 
the stellar surface.
Although this complicated evolution of the convective regions is 
different from that reported in SST80b, this difference can be 
attributable to high sensitivity of the evolution on adopted 
initial deuterium abundance.
We discuss this in greater details in Appendix~\ref{ssec:dab}.

Since the energy generation by D burning is very sensitive 
to the temperature, temperature remains constant at 
$\simeq 1.5 \times 10^{6}$K during the active deuterium burning 
(Fig.~\ref{fig:str_fdtmax_1em5}, lower panel).
This is the so-called thermostat effect (e.g., Stahler 1988).
If this works,
$M_{\ast}/R_{\ast} \propto T \simeq const.$ from equation 
(\ref{eq:t_typ}):
the protostellar radius $R_{\ast}$ increases as $\propto M_{\ast}$.
This linear increase in $R_{\ast}$ with $M_{\ast}$ can be observed 
in the range $0.7M_{\odot} \lesssim M_{\ast} \lesssim 1M_{\odot}$
(Fig.~\ref{fig:str_fdtmax_1em5}, upper panel).

When the deuterium concentration in the convective region
drops to a few percent, the thermostat effect ceases to work.
The maximum temperature starts increasing again and 
the radial expansion slows down after that.
Owing to the high temperature, opacity decreases 
in the deep interior with increasing stellar mass.
This allows efficient radiative energy transport and 
renders a large portion of the interior radiative.
The radiative core gradually extends to the outer layer,  
and the convective layer moves outward.
Since fresh deuterium is no longer supplied to the radiative interior 
by the mixing, deuterium is soon exhausted there. 
The expansion of the radiative interior occurs in a later phase 
in the calculation of PS91.
This difference can be attributed to different initial models: 
whereas PS91 assumed a fully convective star of $1~M_{\odot}$
as an initial model, we started calculation 
from the initial mass $M_{\ast,0} = 0.01~M_{\odot}$.
At $M_{\ast} = 1~M_{\odot}$, the protostar is not fully convective in our case.
The outer layers of $M > 0.25~M_{\odot}$ are convective, 
where the D burning occurs at the bottom, while 
the remaining inner core is radiative at this moment.
We confirmed that, starting from the same initial model, 
the evolution becomes similar to that of PS91 
as discussed in Appendix \ref{ssec:psini}.

Because of the early ignition of deuterium burning, 
entropy and luminosity profiles in this phase look rather different from
those in the high $\dot{M}_\ast$ case 
(see Fig.~\ref{fig:slpf_1em3} and \ref{fig:slpf_1em5}).
For easier comparison, let us see the case without deuterium burning, 
where the similar profiles are in fact reproduced 
(Fig.~\ref{fig:slpf_1em5_nod}).
However, one important difference exists, the spiky structure in the 
entropy profiles near the surface.
This shows significant entropy loss near the surface 
in the low $\dot{M}_\ast$ case, which leads to the low entropy within the star.
The reason can be seen clearly in Figure \ref{fig:t_comp_1em5}, where 
the comparison between the local accretion timescale $t_{\rm acc,s}$ and 
cooling timescale $t_{\rm cool,s}$ (eq.s \ref{eq:taccs} and \ref{eq:tcools})
is presented.
In contrast to the the case of high $\dot{M}_\ast$, 
in most of the settling layer, $t_{\rm acc,s}$ is shorter
than $t_{\rm cool,s}$: accreted matter loses entropy 
radiatively near the surface before settling into the adiabatic interior.

\paragraph{Swelling Phase}

In the short period of $3~M_{\odot} \lesssim M_{\ast} \lesssim 4~M_{\odot}$, 
the stellar radius jumps up by a factor of two. 
At the same time, the whole interior becomes radiative 
(Fig.~\ref{fig:str_fdtmax_1em5}, upper). 
Although deuterium burning continues near the surface, 
it does not drive convection any more.

The cause of this swelling has been attributed to the shell burning 
of deuterium, which also occurs in our calculation (Palla \& Stahler 1990).
However, we conclude that the main driver of the swelling is 
not the D shell burning, but the outward entropy transportation 
as in the case of the high accretion rate.
The reasons are as follows.
First, evolutionary features of entropy and luminosity profiles 
in this phase are similar to those in the swelling
phase in the high $\dot{M}_\ast$ case 
(run MD3; cf. Figs.\ref{fig:slpf_1em3}, \ref{fig:slpf_1em5}):
in both cases, entropy decreases in the deep interior 
and increases significantly near the stellar surface; 
simultaneously, the peak of the luminosity distribution moves to the surface
as the luminosity wave.
Second, even in the run without deuterium burning (run MD5-noD),
this swelling occurs as shown in Figure \ref{fig:str_1em5_nod}: 
the protostar swells up at $M_{\ast} \simeq 3~M_{\odot}$.

The interior luminosity $L_{\ast}$ jumps up at 
$M_{\ast} \simeq 0.7~M_{\odot}$ with the arrival 
of the convective region at the surface 
as a result of high efficiency of convective heat transfer 
(Fig.~\ref{fig:lum_enuc_tsc_1em5}, upper panel).
Although this rise in $L_{\ast}$ decreases $t_{\rm KH}$,  
$t_{\rm KH}$ remains much longer than $t_{\rm acc}$ 
(Fig.~\ref{fig:lum_enuc_tsc_1em5}, lower panel).
It is not until another sudden rise in $L_{\ast}$ at 
$\simeq 3-4 M_{\odot}$ due to the arrival of the luminosity wave 
that $t_{\rm KH}$ becomes as short as $t_{\rm acc}$. 
The maximum radius is attained at this epoch since the swelling is 
also caused by the luminosity wave, as seen above.
Then, the analytical argument leading to equation (\ref{eq:m_rmax}) 
also holds in this case:
for $\dot{M}_\ast = 10^{-5}~M_{\odot}/{\rm yr}$, this leads to 
the maximum radius at $M_{\rm *, rmax} \simeq 3.9~M_{\odot}$, 
which agrees with our numerical result.

Simultaneous returning to the radiative structure 
with the swelling has been explained by \citet{PS90}.
The luminosity from the steady deuterium burning
is comparable to the accretion luminosity by chance:
\begin{equation}
L_{\rm D, st} = \dot{M}_\ast \delta_{\rm D} \sim
               \dot{M}_\ast \frac{G M_{\ast}}{R_{\ast}} = L_{\rm acc} .
\label{eq:ldst_lacc}
\end{equation}
The epoch of the radial turn-around obeys equation
(\ref{eq:m_rmax}), which is derived by equating
$t_{\rm acc}$ and $t_{\rm KH,lmax} = G M_*^2/R_* L_{\rm max}$.
This means that $L_{\rm max}$ becomes comparable to $L_{\rm acc}$ 
and thus to $L_{\rm D,st}$ in the swelling phase.
Hence, the radiative luminosity now has a capacity to transport
the energy generated by deuterium burning: 
$L_{\rm max} \sim L_{\rm D,st}$.  
The deuterium burning no longer gives rise to any convective layer 
after the swelling.

\paragraph{Later Phases}

Subsequently, the protostar enters the KH contraction phase.
The evolution thereafter is similar to the previous high-$\dot{M}_\ast$ case. 
The maximum temperature within the star increases with the contraction, and
hydrogen burning begins at $M_{\ast} \simeq 7~M_{\odot}$.

\begin{figure}[t]
  \begin{center}
\epsfig{ file=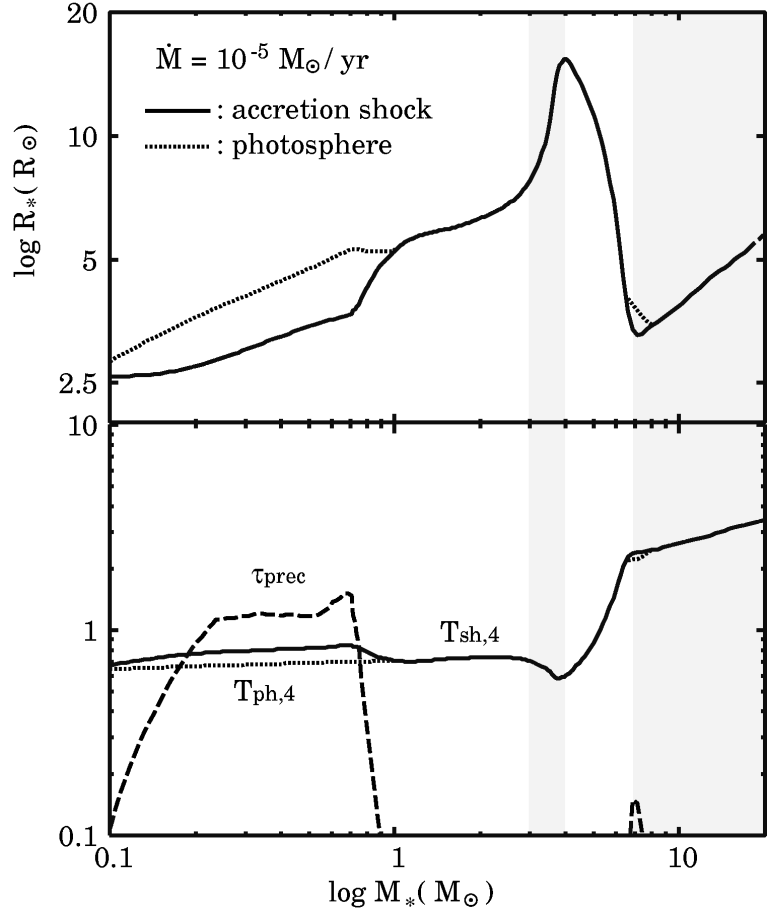,
         angle=0,
         width=4in}
\caption{Same as Fig.~\ref{fig:prec_1em3} but for the lower
accretion rate of $\dot{M}_\ast = 10^{-5}~M_{\odot}/{\rm yr}$ (run MD5). 
The shaded background shows the four characteristic phases,
as in Fig.~\ref{fig:str_fdtmax_1em5}. 
}
\label{fig:prec_1em5}
  \end{center}
\end{figure}

\paragraph{Evolution of the Radiative Precursor}

Figure \ref{fig:prec_1em5} shows
that the accretion flow is optically thin except 
in the early phase of $M_{\ast} < 1~M_{\odot}$
at the low accretion rate of $10^{-5}~M_{\odot}/{\rm yr}$.
With the expansion at the active deuterium burning, 
the stellar surface emerges above 
the photosphere at $M_{\ast} \simeq 1~M_{\odot}$ 
and the radiative precursor disappears thereafter. 
Without the precursor, 
the stellar surface is directly visible, 
with effective temperature rising remarkably 
in the KH contraction phase, i.e., for 
$M_{\ast} > 4 M_{\odot}$.
This is in stark contrast to the case of 
high accretion rate $10^{-3}~M_{\odot}/{\rm yr}$, 
where the effective temperature remains several thousand K
throughout the evolution (Fig.~\ref{fig:prec_1em3}).

\subsection{Dependence on Mass Accretion Rates}
\label{ssec:md_dep}

\begin{figure}[t]
  \begin{center}
\epsfig{ file=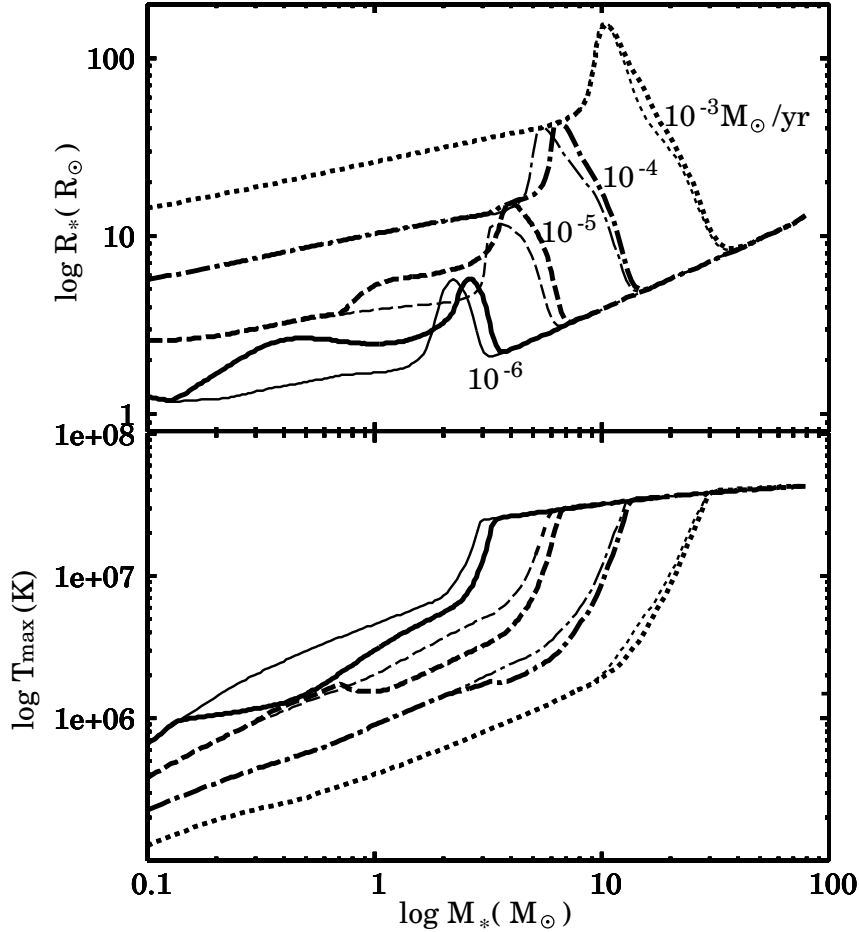,
         angle=0,
         width=4.5in}
\caption{
Evolution of the protostellar radii (upper panel) 
and the maximum temperatures within the stars (lower panel) 
for different mass accretion rates. 
The cases of $\dot{M}_\ast=10^{-6}$ (run MD6; solid), 
$10^{-5}$ (MD5; dashed), $10^{-4}$ (MD4; dot-dashed), 
and $10^{-3} M_{\odot}/{\rm yr}$ (MD3; dotted line) are depicted.
Also shown by the thin lines are the runs without 
deuterium burning (``noD'' runs) for the same accretion rates.
In both panels, all the curves finally converge to single lines, 
which is the relations for the main-sequence stars.
}
\label{fig:m_rtmax}
  \end{center}
\end{figure}
\begin{figure}[t]
\begin{center}
\epsfig{ file=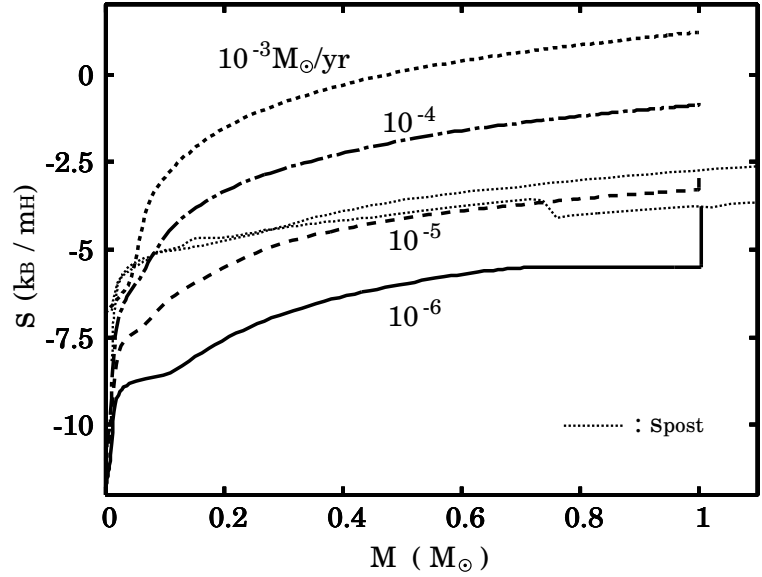,
         angle=0,
         width=4in}
\caption{
Distributions of the specific entropy for the case without deuterium
(``no-D'' runs in Table 1) at the epoch of $M_{\ast} = 1~M_{\odot}$ 
for different accretion rates. 
Histories of the post-shock values are also plotted by the dotted lines
for the cases with $\dot{M}_\ast = 10^{-6}$ and 
$10^{-5}~M_{\odot}/{\rm yr}$.
For the cases with $\dot{M}_\ast = 10^{-4}$ and $10^{-3}~M_{\odot}/{\rm yr}$, 
those curves completely overlap with the entropy distributions
at $M_{\ast} = 1~M_{\odot}$.
}
\label{fig:s_1msun_nod}
\end{center}
\end{figure}
\begin{figure}[h]
  \begin{center}
\epsfig{ file=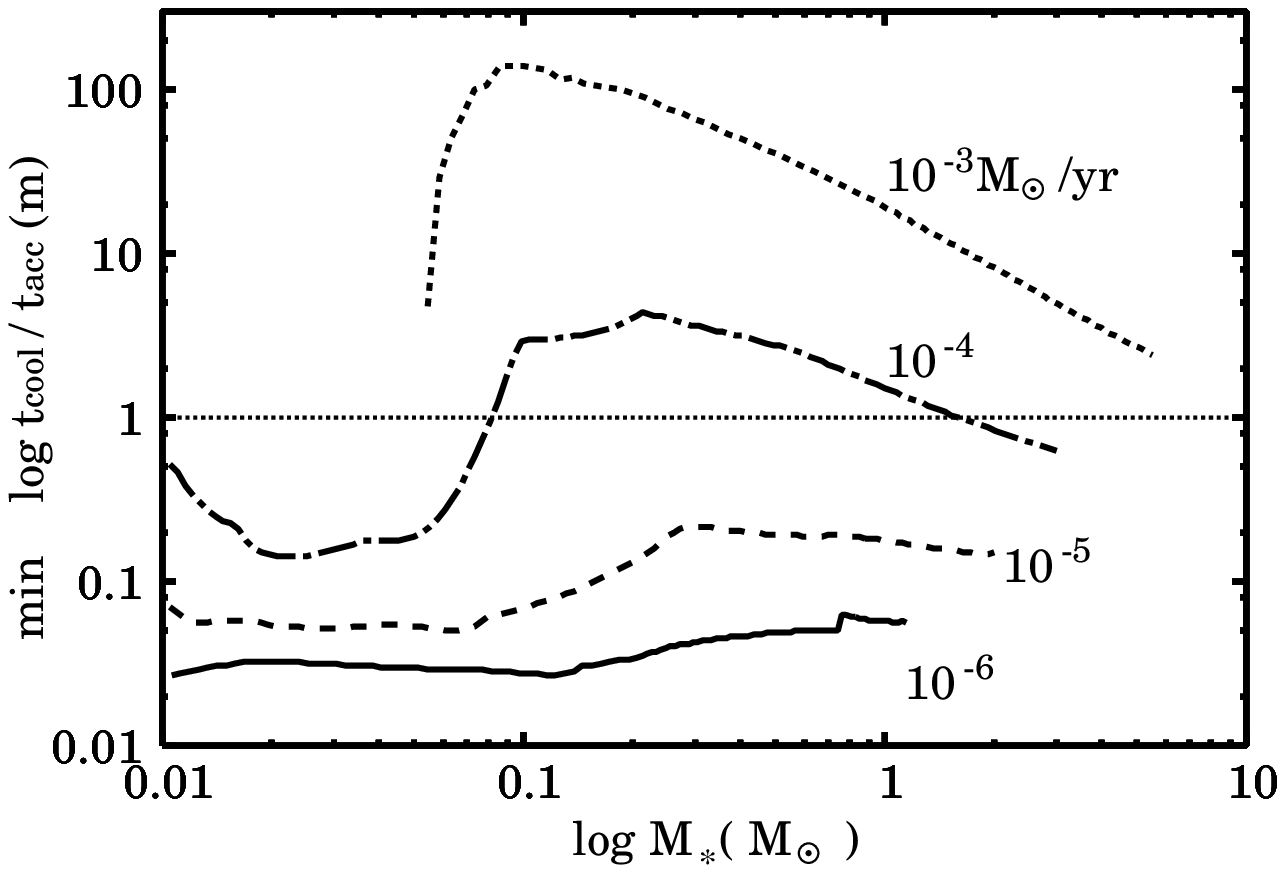,
         angle=0,
         width=4in}
\caption{ 
Evolution of the minimum value 
of the ratio $t_{\rm cool,s}(m)/t_{\rm acc,s}(m)$ 
in the surface settling layer.
The no-D cases are shown for 
$\dot{M}_\ast=10^{-6}$ (solid), $10^{-5}$ (dashed), 
$10^{-4}$ (dot-dashed), and $10^{-3}~M_{\odot}/{\rm yr}$ (dotted).
For the ratio $t_{\rm cool,s}(m)/t_{\rm acc,s}(m)$ exceeding unity, 
the accreted material settles into the interior without losing 
its post-shock entropy radiatively. 
}
\label{fig:r_tcl_tacc}
  \end{center}
\end{figure}

Here, we summarize the effects of the different accretion rates 
on the protostellar evolution.
The mass-radius relations of protostars with 
accretion rates $\dot{M}_\ast=10^{-6}$, 
$10^{-5}$, $10^{-4}$, and $10^{-3} M_{\odot}/{\rm yr}$
are shown in the upper panel of Figure \ref{fig:m_rtmax}.

First, the protostellar radius is larger for the higher accretion
rate at the same protostellar mass, 
as discussed in \S~\ref{ssec:md_1em3}.  
For example, at $M_{\ast} = 1 M_{\odot}$, the radius
$R_{\ast} \simeq 25 R_{\odot}$ 
for $10^{-3} M_{\odot}/{\rm yr}$, while it is only 
$2.5 R_{\odot}$ for $10^{-6} M_{\odot}/{\rm yr}$.
For easier comparison, the mass-radius relations in the no-deuterium 
cases are also shown by thin lines in Figure \ref{fig:m_rtmax}.
The deuterium burning enhances the stellar radius,  
in particular, in low $\dot{M}_\ast$ cases
(see in this section below).
Nevertheless, the trend of larger $R_{\ast}$ for higher $\dot{M}_\ast$ 
remains valid even without the deuterium burning.
For simplicity, we do not include the effect of 
deuterium burning in the following argument.
The origin of this trend is higher specific entropy for 
higher accretion rate.
Recall that the stellar radius is larger with the high
entropy in the stellar interior as shown by equation (\ref{eq:r_srel}).
The entropy distributions at $M_{\ast} = 1 M_{\odot}$ 
for the no-D runs are shown in Figure \ref{fig:s_1msun_nod}, 
which indeed demonstrates that 
the entropy is higher for the higher $\dot{M}_\ast$ cases.
The interior entropy is set in the post-shock settling layer 
where radiative cooling can reduce the entropy from 
the initial post-shock value 
as observed in spiky structures near the surface 
(e.g., the case of $\dot{M}_\ast \lesssim 10^{-5}~M_{\odot}/{\rm yr}$ 
in Fig.~\ref{fig:s_1msun_nod}). 
For efficient radiative cooling,
the local cooling time $t_{\rm cool, s}$ in the settling layer
must be shorter than the local accretion time $t_{\rm acc,s}$,
as explained in \S~\ref{ssec:md_1em3} and \ref{ssec:md_1em5}.
Figure \ref{fig:r_tcl_tacc} shows that the ratio 
$t_{\rm cool,s}/t_{\rm acc,s}$ 
is smaller for lower $\dot{M}_\ast$ and becomes less than unity
for $\lesssim 10^{-5}~M_{\odot}/{\rm yr}$.
Thus the accreted gas cools radiatively in the settling layer 
in low $\dot{M}_\ast$ ($\lesssim 10^{-5} M_{\odot}/{\rm yr}$) cases, 
while in the higher $\dot{M}_\ast$ cases, the gas is embedded into the stellar 
interior without losing the post-shock entropy.
The difference in entropy among high 
$\dot{M}_\ast$ ($\gtrsim 10^{-4} M_{\odot}/{\rm yr}$) cases comes from 
the higher post-shock entropy for higher $\dot{M}_\ast$.
In those cases, the protostar is enshrouded with the 
radiative precursor, whose optical depth is 
larger for the higher accretion rate 
(c.f.,Figs.\ref{fig:prec_1em3} and \ref{fig:prec_1em5}, bottom).
Consequently, more heat is trapped in the precursor, 
which leads to the higher post-shock entropy.

Figure \ref{fig:m_rtmax} shows not only that 
not only the swelling of the stellar radius occurs even in the ``no-D'' runs,
but also that their timings are almost the same as in the cases with D.
This supports our view that the cause of the swelling is 
outward entropy transport by the luminosity wave
rather than the deuterium shell-burning 
(\S~\ref{ssec:md_1em3} and \ref{ssec:md_1em5}).
Also, in all cases, the epoch of the swelling obeys equation 
(\ref{eq:m_rmax}), which is derived by equating
the accretion time $t_{\rm acc}$ and the KH time $t_{\rm KH,lmax}$.

Another clear tendency is that
the deuterium burning begins at higher stellar mass for 
higher accretion rate.
Similarly, the onset of hydrogen burning and subsequent arrival at the ZAMS 
are postponed until higher protostellar mass for the higher accretion rate.
For example, the ZAMS arrival is at $M_{\ast} \simeq 4 M_{\odot}$ 
(40$M_{\odot}$) for $10^{-6}$ ($10^{-3}$, respectively) 
$M_{\odot}/{\rm yr}$.
These delayed nuclear ignitions reflect the lower temperature 
within the protostar for the higher $\dot{M}_\ast$ at the same $M_{\ast}$ 
(Fig.~\ref{fig:m_rtmax}, lower panel).
Since $R_{\ast}$ is larger with higher $\dot{M}_\ast$ at a given $M_{\ast}$,
typical temperature is lower within the star.
Substituting numerical values in equation (\ref{eq:t_typ}), we obtain
\begin{equation}
T_{\rm max} \sim 10^7~{\rm K} 
                 \left( \frac{M_{\ast}}{M_{\odot}}   \right)
                 \left( \frac{R_{\ast}}{R_{\odot}}   \right)^{-1} ,
\label{eq:t_max_aly}
\end{equation}
which is a good approximation for our numerical results.

Finally, we consider the significance of deuterium burning 
for protostellar evolution. Comparing  with ``no-D'' runs,
we find that the effect of deuterium burning not only appears
later, but also becomes weaker with increasing accretion rates.
At the low accretion rates
$\dot{M}_\ast \leq 10^{-5}~M_{\odot}/{\rm yr}$, 
convection spreads into a wide portion of the star owing to the vigorous 
deuterium burning. 
In addition, the thermostat effect works and the stellar radius increases 
linearly during this period. At the higher accretion rates 
$\dot{M}_\ast \geq 10^{-4}~M_{\odot}/{\rm yr}$, on the other hand, 
deuterium burning affects the protostellar evolution only slightly. 
Convective regions do not appear even after the beginning 
of deuterium burning for $\dot{M}_\ast = 10^{-3} M_{\odot}/{\rm yr}$ 
(\S~\ref{ssec:md_1em3}).

A key quantity for understanding this variation is the
maximum radiative luminosity at the ignition of deuterium burning.
The maximum radiative luminosity within the star $L_{\rm max}$
at deuterium ignition can be evaluated by 
eliminating $R_{\ast}$ and $M_{\ast}$ in equation (\ref{eq:l_max_a})
with (\ref{eq:r_sps86}), (\ref{eq:t_max_aly}), and setting
$T_{\rm max} \sim 10^6$~K: 
\begin{equation}
L_{\rm max} \sim 10^3~L_{\odot} 
                    \left( \frac{\dot{M}_\ast}{10^{-3}~M_{\odot}/{\rm yr}} 
                    \right)^{2.8} .  
\end{equation}
On the other hand, the luminosity from deuterium burning is 
(eq. \ref{eq:l_dst})
$L_{\rm D,st} \sim 10^3~L_{\odot}~(\dot{M}_\ast/10^{-3}~M_{\odot}/{\rm yr})$. 
For accretion rate $\dot{M}_\ast \geq 10^{-3}~M_{\odot}/{\rm yr}$, 
$L_{\rm max}> L_{\rm max, D}$:
radiation is able to carry away all the entropy generated by 
the deuterium burning. 
Therefore, the protostellar interior remains radiative 
in the high $\dot{M}_\ast$ cases (e.g., run MD3, see \S~\ref{ssec:md_1em3}), 
while convection is excited for the lower $\dot{M}_\ast$ 
(e.g., run MD5, see \S~\ref{ssec:md_1em5}).

\subsection{Comparison with Primordial Star Formation}
\label{ssec:metal}

\begin{figure}[t]
  \begin{center}
\epsfig{ file=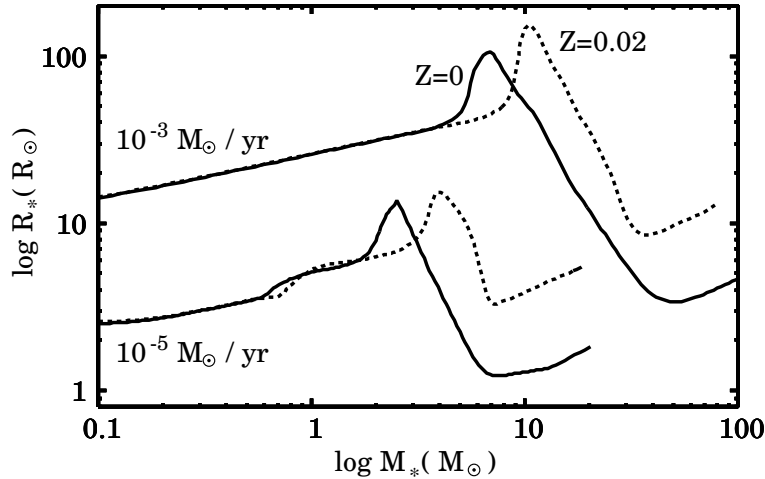,
         angle=0,
         width=4in}
\caption{The protostellar mass-radius relations for different
metallicities.
The cases of the solar (dotted; runs MD5 and MD3) 
and zero (solid; MD5-z0 and MD3-z0) metallicities 
are presented 
for the accretion rates $\dot{M}_\ast=10^{-5}$ and $10^{-3} 
M_{\odot}/{\rm yr}$.
}
\label{fig:m_r_z}
  \end{center}
\end{figure}

Besides the massive star formation in the local universe,  
high accretion rates have been invoked in the primordial 
star formation in the early universe.
Without important metal coolants, the temperature 
remains at several hundred K in the primordial gas 
\citep[e.g.,][]{SZ67, MST67, PSS83}.
The accretion rate, which is given by  
\begin{equation}
\dot{M}_\ast \simeq \frac{c_{\rm s}^3}{G} 
= 10^{-3} M_{\sun}/{\rm yr}
\left( \frac{T}{500{\rm K}} \right)^{3/2},
\end{equation}
where $c_{\rm s}$ and $T$ are the sound speed and temperature, respectively, 
in star-forming clouds, 
becomes as high as $10^{-3}M_{\odot}/{\rm yr}$ even without turbulence 
(SPS86).
As a result of similar accretion rates, 
protostellar evolution of the primordial stars well resembles  
with those of the high $\dot{M}_\ast$ cases studied in this paper  
(SPS86 ; Omukai \& Palla 2001,2003).
The mass-radius relations for the primordial ($Z=0$) protostars are 
presented in Figure \ref{fig:m_r_z}, along with their solar-metallicity 
($Z=0.02$) counterparts for comparison.
This indicates that basic evolutionary features are 
similar despite differences in metallicity:
as in the present-day protostars, 
primordial protostars pass through the same four characteristic
phases (see \S~\ref{ssec:md_1em3} and \ref{ssec:md_1em5}).
The following two differences, however, still exist.
First, the swelling of the protostar and subsequent 
transition to the KH contraction
occur earlier in the $Z=0$ cases.
For example, for $\dot{M}_\ast=10^{-3}M_{\sun}/{\rm yr}$, 
the maximum radius is attained and the KH contraction 
begins at $10M_{\sun}$ in the $Z=0.02$ case, while it is $7M_{\sun}$
in the $Z=0$ case.
Recall that the transition to those phases is caused by 
the decrease in opacity with increasing $M_{\ast}$ (\S~\ref{ssec:md_1em3}).
Owing to the lack of heavy elements, the opacity is lower 
in the primordial gas at the same density and temperature.
Efficient radiative heat transport begins earlier, which 
results in the earlier transitions in the evolutionary phases.
Second, the stellar radius in the main-sequence phase 
is smaller as a result of higher interior temperature 
($\simeq 10^{8}$K) for the primordial stars.
Due to the lack of C and N, sufficient energy production 
by the CN cycle is not available at a temperature similar to that 
of the solar metallicity case ($\simeq 10^{7}$K).
The KH contraction is not halted until 
the temperature reaches $\simeq 10^{8}$K, where 
a small amount of carbon produced by the He burning
enables the operation of the CN cycle
\citep{EC71, OP03}.
Since more contraction is needed, the arrival at the ZAMS can be later 
in the $Z=0$ case, despite the earlier beginning of the KH contraction:
for example, it is $30M_{\odot}$ ($50M_{\odot}$) in the $Z_{\odot}$ 
($Z=0$, respectively) case.

\subsection{Radiation Pressure Barrier on the Accreting Gas Envelope}
\label{ssec:edd}

\begin{figure}[t]
  \begin{center}
\epsfig{ file=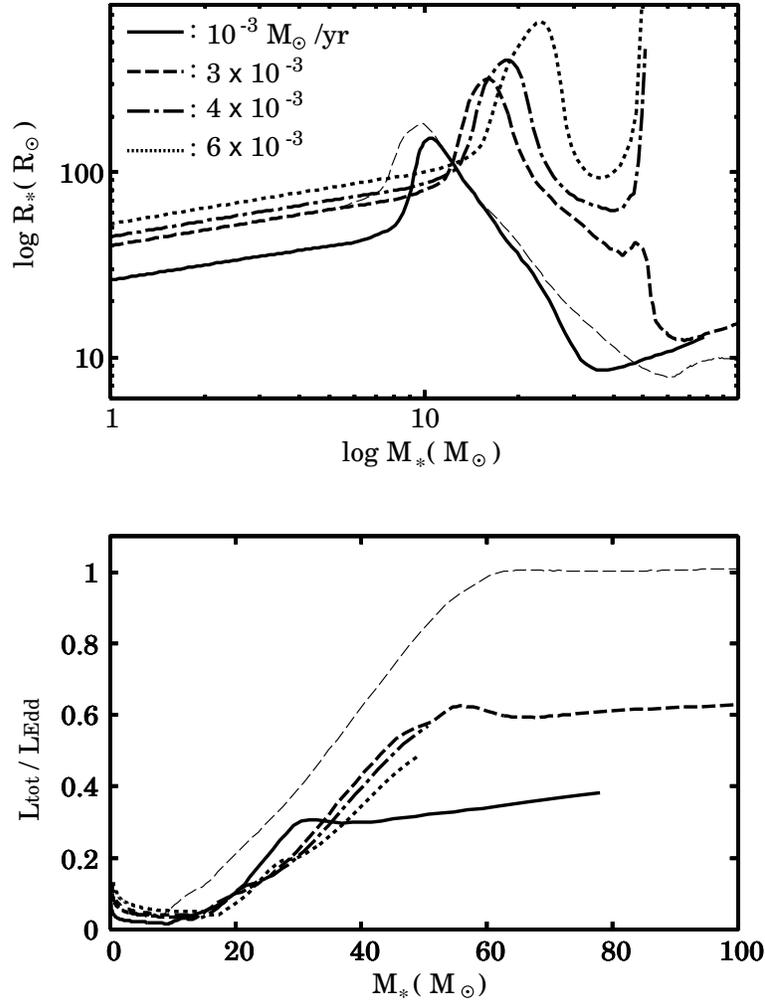,
         angle=0,
         width=4in}
\caption{ 
{\it Upper panel} :
Evolution of the protostellar radii for cases
at the very high accretion rates
of $10^{-3}$ (solid, run MD3), $3 \times 10^{-3}$ 
(dashed, MD3x3), $4 \times 10^{-3}$ 
(dot-dashed, MD4x3), and $6 \times 10^{-3}~M_{\odot}/{\rm yr}$ 
(dotted, MD6x3).
For comparison, we also plot the case of a
primordial ($Z = 0$) protostar with $3 \times 10^{-3}~M_{\odot}/{\rm yr}$ 
(thin dashed, MD3x3-z0).
{\it Lower panel} : Evolution of a ratio between the total luminosity
of protostars and Eddington luminosity for the same cases. 
}
\label{fig:m_r_blowup}
  \end{center}
\end{figure}         

The higher the mass accretion rate, the more massive the 
protostar grows before the arrival at the ZAMS phase
(\S~\ref{ssec:md_1em3} and \ref{ssec:md_dep}).
However, this trend does not continue unlimitedly: 
for a rate exceeding a threshold value, stars have an upper mass 
limit set by the radiation pressure onto the radiative 
precursor. 

The evolution of the protostellar radius is shown in 
Figure \ref{fig:m_r_blowup} (upper panel) for cases with 
very high accretion rates $>10^{-3}~M_{\odot}/{\rm yr}$.
Up to a certain moment in the KH contraction phase, 
the evolution in all cases proceeds in a qualitatively 
similar way 
although the radius is larger and the onset of swelling 
and the KH contraction are later for higher $\dot{M}_\ast$ cases.
However, for accretion rates 
$\dot{M}_\ast > 3 \times 10^{-3} M_{\odot}/{\rm yr}$, 
the contraction is halted at $\simeq 40M_{\sun}$, followed by
abrupt expansion at $\simeq 50M_{\sun}$.
In such cases, further steady accretion onto the star is not possible
owing to strong radiative pressure exerted onto the radiative precursor.
This phenomenon has been found for the primordial star formation 
by \citet{OP01, OP03}.
They found that abrupt expansion occurs when the total
luminosity from a protostar,
$L_{\rm tot} = L_{\ast} + L_{\rm acc}
=L_{\ast} + G M_{\ast} \dot{M}_{\ast}/R_{\ast}$, 
becomes close to the Eddington luminosity 
$L_{\rm Edd}$.
The strong radiation pressure decelerates the accretion flow and then 
reduces the ram pressure onto the stellar surface. 
With reduced inward ram pressure,  
an outward thrust by the interior radiation 
blows a thin surface layer away.

The critical accretion rate $\dot{M}_{\ast}$ is defined as the rate above 
which the abrupt stellar expansion occurs in the KH contraction phase.
By using the condition that $L_{\rm tot}$ reaches $L_{\rm Edd}$ before 
the protostar reaches the ZAMS, \citep{OP03} derived the 
critical rate as follows:
\begin{equation}
L_{\rm ZAMS} + \frac{G M_{\rm ZAMS} \dot{M}_{\rm \ast, cr}}{R_{\rm ZAMS}} 
= L_{\rm Edd},
\label{eq:lcr_pm}
\end{equation}
that is,
\begin{equation}
\dot{M}_{\rm \ast, cr} = \frac{4 \pi c R_{\rm ZAMS}}{\kappa_{\rm esc}}
                   \left( 1 - \frac{L_{\rm ZAMS}}{L_{\rm Edd}}  \right)
                   ,
\label{eq:mcr_pm}
\end{equation}
where $\kappa_{\rm esc}$ is the electron-scattering opacity, 
and the suffix ``ZAMS'' means quantities of the ZAMS stars. 
Although the critical rate $\dot{M}_{\rm \ast, cr}$ 
by equation (\ref{eq:mcr_pm}) depends on 
the stellar mass, the dependence is very weak and 
$\dot{M}_{\rm \ast, cr} \simeq 4 \times 10^{-3} M_{\odot}/{\rm yr}$ 
for the primordial stars.
Since the ZAMS radius increases with metallicity 
(see \S~\ref{ssec:metal}), 
$\dot{M}_{\rm \ast, cr}$ is expected to increase with metallicity.
For example, for $Z=10^{-4}$, 
$\dot{M}_{\rm \ast, cr}=9 \times 10^{-3} M_{\odot}/{\rm yr}$ by 
equation (\ref{eq:mcr_pm}), and the evolutionary calculation 
demonstrated that abrupt stellar expansion occurs in fact 
for higher accretion rates \citep{OP03}.
Our calculations, however, indicate that the critical accretion rate in 
the solar metallicity case is $\simeq 3 \times 10^{-3} 
M_{\odot}/{\rm yr}$, slightly smaller than that of primordial protostars.
This apparent contradiction comes from the fact that, 
in the solar metallicity case, opacity in the accreting envelope is 
higher than that of the electron scattering one $\kappa_{\rm es}$ 
postulated in the above derivation for the critical accretion rate.
In fact, the violent stellar expansion is found to take place
with a total luminosity of about a half of the Eddington luminosity, 
as shown in the lower panel of Figure \ref{fig:m_r_blowup}.
A marginal case is Run MD3x3, where the contraction is 
about to be reversed, but the star barely reaches 
the ZAMS after some loitering. 
In this case, the Eddington ratio 
$f_{\rm Edd}=L_{\rm tot}/L_{\rm Edd}$ slightly exceeds 
0.5 at $M_{\ast} \simeq 45~M_{\odot}$.
For comparison, 
the case of the primordial star is shown for the same accretion rate, 
$\dot{M}_\ast = 3 \times 10^{-3}~M_{\odot}/{\rm yr}$, 
which is slightly lower than the critical accretion rate (run MD3x3-z0).
Because of the earlier transition to the KH contraction phase, 
the Eddington ratio also begins to increase
earlier than the present-day counterpart (run MD3x3).
Despite high luminosity almost reaching the Eddington limit, 
the KH contraction is not reversed and the protostar can grow 
further.
Taking into account the higher opacity in the envelope, 
we add the Eddington factor of $0.5$ to 
equation (\ref{eq:lcr_pm}):
\begin{equation}
L_{\rm ZAMS} + \frac{G M_{\rm ZAMS} \dot{M}_{\rm \ast, cr}}{R_{\rm ZAMS}}
\simeq 0.5 L_{\rm Edd}
\label{eq:lcr_present}
\end{equation}
for present-day protostars. 
The critical accretion rate becomes 
$\dot{M}_{\rm \ast, cr} \simeq 3 \times 10^{-3} M_{\odot}/{\rm yr}$,
which agrees well with our numerical results. 
Although this critical rate is a function of the stellar mass, 
its dependence is weak as in the primordial case.

Our results suggest an upper
mass limit of PMS stars at $\simeq 60~M_\odot$, 
which corresponds to the mass where the protostar reaches 
the ZAMS at the critical accretion rate of 
$M_{\rm cr} \sim 3 \times 10^{-3}~M_\odot/{\rm yr}$.
More massive PMS stars cannot be formed by steady
mass accretion. With higher accretion rates, the steady 
accretion is halted by the abrupt expansion at
$M_\ast \simeq 50~M_\odot$. 
What happens after the abrupt expansion is speculated
upon in the next section.

\section{Upper Limit on the Stellar Mass by Radiation Feedback}
\label{sec:stopacc}

\begin{figure}[t]
  \begin{center}
\epsfig{ file=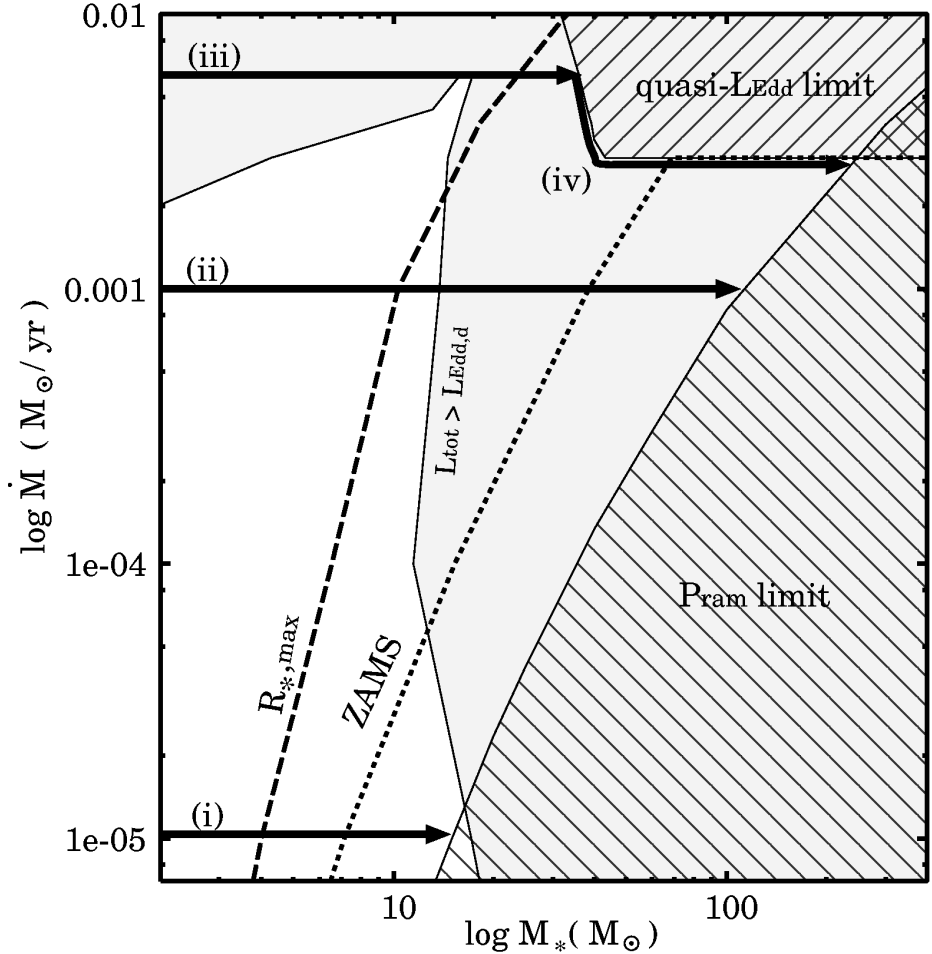,
         angle=0,
         width=5in}
\caption{ 
Limits on the stellar mass for each accretion rate by various
feedback effects. 
The three horizontal arrows represent 
protostellar evolution at the accretion rate of (i)
$\dot{M}_\ast = 10^{-5}$ (run MD5), (ii) $10^{-3}$ (run MD3), 
and (iii) $6 \times 10^{-3}~M_{\odot}/{\rm yr}$ (run MD6x3).
Also shown are the stellar masses where the protostellar radii
reach the maximum (dashed line) and where the protostars arrive
at the ZAMS (dotted line).
The region shaded by light gray represents where the
total luminosity of the protostar, 
$L_{\rm tot} = L_{\ast} + L_{\rm acc}$ exceeds the dust Eddington
luminosity $L_{\rm Edd,d}$, given by equation (\ref{eq:eddd}).
The lower-right hatched region represents where the ram
pressure of the free-fall flow is insufficient to overcome 
the radiation pressure barrier at the dust destruction front.
The upper-right hatched region denotes the similar
prohibited region but for the radiation pressure acting on
a gas envelopes just around the protostar. 
The extra arrow (iv) indicates a possible path along which
the protostar will evolve after the gas envelope is disrupted
by the radiation pressure (see text).
}
\label{fig:wc}
  \end{center}
\end{figure}         

As long as the central star is not so massive and
its feedback on the accreting envelope is weak, 
the accretion proceeds at a rate set by the condition of 
the parent star-forming core. 
Our assumption concerning the accretion at a pre-determined rate 
remains valid.
However, when the central star grows massive enough and
its feedback, e.g. radiative pressure on 
the dusty outer envelope, thermal pressure in the HII region, etc., 
can no longer be neglected,
the accretion rate is now regulated by the stellar feedback. 
The mass accretion is eventually terminated  
and the final stellar mass is set at this moment.

Among possible types of feedback, the radiation pressure 
on the outer dusty envelope is considered to be the most important 
in terminating the protostellar accretion (e.g., WC87).
Since no such outer envelope has been included in our calculation, 
we evaluate here the epoch of accretion termination  
by stellar feedback with the assumption that the accretion rate 
is roughly constant up to this moment.

Most of the stellar radiation, emitted in the optical or UV 
range, is absorbed in a thin innermost layer 
of the dusty envelope, i.e., the dust destruction front, 
and the outward momentum of photons 
is transfered to the flow (see Fig.~\ref{fig:prost_schem}).
Overcoming this radiation pressure barrier is a necessary 
condition for continuing accretion.
The condition that the outward radiation pressure falls below
the inward ram pressure of the accretion flow is expressed as
\begin{equation}
\frac{L_{\rm tot}}{4 \pi R_{\rm d}^2 c} < \rho u^{2},
\end{equation}
where $R_{\rm d}$ is the radius of the dust destruction front, 
and $\rho$ and $u$ are the flow density and velocity there.
The limit on the stellar mass 
by this condition is shown in Figure \ref{fig:wc}
as a function of the mass accretion rate (i.e, ``$P_{\rm ram}$ limit'').
Here, we have assumed that the flow is in the free-fall.
In Figure \ref{fig:wc}, the epochs of the maximum swelling and the arrival 
at the ZAMS are also presented from our results. 
Except in cases with very high accretion rates 
($> 3 \times 10^{-3}~M_{\odot}/{\rm yr}$; \S~\ref{ssec:edd}),
the radiation pressure limit is attained 
after the stars reach the ZAMS.
This verifies previous authors' treatment in assuming 
that the central star is already in the ZAMS phase when 
the feedback becomes important (e.g., WC87). 

In reality, the accretion flow is decelerated
before reaching the dust destruction front. 
The radiation absorbed at the dust destruction
front is re-emitted in infrared wavelengths and 
this re-emitted radiation imparts pressure onto 
the outer flow.
This radiation pressure exceeds the gravitational pull of the star 
if the luminosity exceeds (McKee \& Ostriker 2007):
\begin{equation}
L_{\rm Edd,d} = \frac{4 \pi c G M_{\ast}}{\kappa_{\rm d, IR}}
\simeq  1600 L_{\odot} \left( \frac{M_{\ast}}{M_{\odot}}  \right), 
\label{eq:eddd}
\end{equation}
which is the ``Eddington luminosity'' for dusty gas, 
calculated by using the dust opacity for the far-infrared light
$\kappa_{\rm d, IR} \simeq 8~{\rm cm^2 / g}$  
instead of the electron-scattering opacity.
The adopted value $8~{\rm cm^2 / g}$ is the Rosseland mean opacity 
at $T \simeq 600$~K, where the opacity takes the maximum value
\citep{Pol94}. 
In the outermost region of the dust cocoon, 
where $T < 600$~K and $\kappa_{\rm d, IR} < 8~{\rm cm^2 / g}$,
the deceleration effect is somewhat milder.
This effect becomes significant when the flow reaches
the inner warm region.
As shown by the light-shaded region in Figure \ref{fig:wc},
the deceleration of dusty flow occurs ($L_{\rm tot} > L_{\rm Edd,d}$) 
in a wide range of $M_{\ast}$ and $\dot{M}_\ast$.
In this case, the upper mass limit by the radiation pressure 
(``$P_{\rm ram}$ limit'' in Fig.~\ref{fig:wc}), which is derived 
by assuming the free-fall flow, becomes lower: the boundary of the 
``$P_{\rm ram}$ limit'' region shifts to the left to some extent.
To evaluate this effect more quantitatively and then
to identify how mass accretion is finally terminated,
further studies will be needed.
It has been shown that the deceleration
effect of infrared light is diminished by reduced dust opacity (WC87),
non-spherical accretion \citep[e.g.,][]{Nk89, YS02},
and other various processes 
(e.g., see McKee \& Ostriker 2007 for a recent review).
Note that even with the reduced dust opacity, the ``$P_{\rm ram}$
limit'' still remains as long as the dusty envelope is thick enough
for optical/UV (not infrared) light from the star. 

As seen in \S~\ref{ssec:edd}, for a very high accretion rate of  
$\dot{M}_\ast > 3 \times 10^{-3}~M_{\odot}/{\rm yr}$, 
the radiation pressure acting on the gas 
envelope terminates the steady mass accretion. 
Figure \ref{fig:wc} shows that 
this limit (``quasi-$L_{\rm Edd}$ limit'') is more severe 
than the radiation pressure limit 
for $\dot{M}_\ast > 3 \times 10^{-3}~M_{\odot}/{\rm yr}$ 
although the deceleration of the dusty flow is already important.
Note that WC87's result on this quasi-Eddington limit 
differs from ours (see their Fig.~5).
This discrepancy comes from their assumption that the forming star is 
already in the ZAMS, which is not valid in that case.
On the other hand, we have consistently calculated the evolution  
until this limit is reached and the star begins abrupt expansion.
For such a high accretion rate, 
what happens when the steady accretion
becomes impossible is not clear in realistic situations.
The accreting envelope may be completely 
blown away and growth of the protostar may be terminated 
at this point.
Without accretion, the protostar quickly relaxes 
to a ZAMS star. 
Another possibility is that since the strong radiation 
force originates in the accretion luminosity itself, 
the accretion may continue in a self-regulated way 
by reducing its rate, for example, 
by an outflow.
This process may proceed in a sporadic way:
the protostar continues to grow alternately repeating 
eruption and accretion.
If this is the case, the protostar evolves along the trajectory (iv)
in Figure \ref{fig:wc}, which runs along the boundary of the 
prohibited region and the maximum stellar mass is 
about $250~M_{\odot}$ set by the radiation pressure feedback 
at the critical accretion rate 
$\dot{M}_{\rm \ast, cr} \simeq 3 \times 10^{-3}~M_{\odot}/{\rm yr}$ 
(Fig.~\ref{fig:wc}).
This is a somewhat higher value than the likely upper mass 
cut-off of the Galactic initial mass function 
$\sim 150~M_{\odot}$ \citep[e.g.,][]{WK04, Fg05}.
This disagreement might be due to the deceleration of the outer 
dusty envelope, which is not included in this analysis.

\section{Observational Signatures of High Accretion Rates}
\label{sec:obs}

\begin{figure}[t]
  \begin{center}
\epsfig{ file=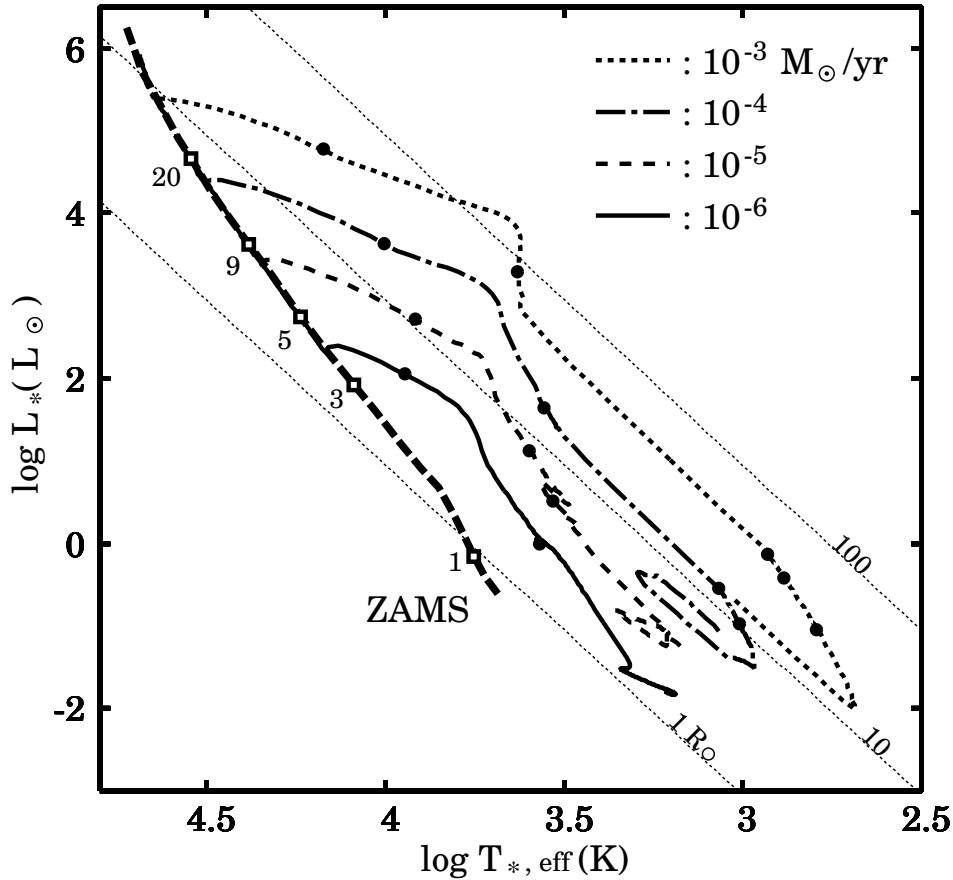,
         angle=0,
         width=5in}
\caption{
The stellar birthlines for different accretion rates. 
The cases for accretion rates $10^{-6}~M_{\odot}/{\rm yr}$ 
(solid; run MD6), $10^{-5}$ (dashed; MD5), $10^{-4}$ (dot-dashed; MD4), 
and $10^{-3}~M_{\odot}/{\rm yr}$ (dotted; MD3) are presented.
Each track shows the evolution from the initial model, 
and filled circles on the track represent points of the stellar
mass $M_{\ast}=$ 1, 3, 5, 9, and 20$M_{\odot}$ 
from the lower right in this order. 
The thick broken line represents the positions for 
zero-age main-sequence from \citet{Sch92}.
The open squares along the line denote the positions for the 
same masses as above.
The dotted straight lines indicate the loci for the 
constant stellar radius of $1 R_{\odot}$,
$10 R_{\odot}$, and $100 R_{\odot}$.  }
\label{fig:hr}
  \end{center}
\end{figure}
\begin{figure}[t]
  \begin{center}
\epsfig{ file=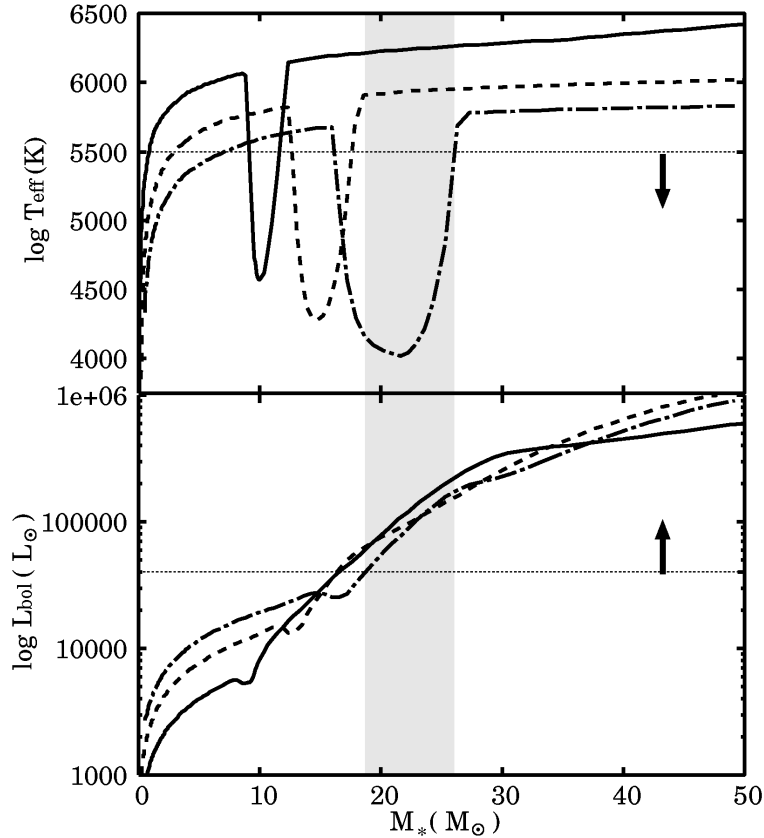,
         angle=0,
         width=4in}
\caption{
Evolution of the effective temperatures and bolometric luminosities 
of protostars with accretion rate $\dot{M}_\ast=10^{-3}$ (solid; run MD3),
$3 \times 10^{-3}$ (dashed; MD3x3), and 
$6 \times 10^{-3} M_{\odot}/{\rm yr}$ (dot-dashed; MD6x3).
The condition required for the protostar in Orion KL region \citep{Mn98}
is indicated by the horizontal lines with arrows;
$L_{\rm tot} \geq 4 \times 10^4~L_{\odot}$ and $T_{\rm eff} < 5500$K.
The mass range satisfying this condition is denoted by the shade 
for the case of $\dot{M}_\ast=6 \times 10^{-3} M_{\odot}/{\rm yr}$. } 
\label{fig:KL}
  \end{center}
\end{figure}

Massive protostars with high accretion rates 
have some characteristic observational features.
In this section, we explore possibilities of identifying them
observationally and of verifying the postulated high accretion rates.
Although protostars still in the main accretion phase are hard to 
observe directly, pre-main sequence (PMS) stars just 
after this phase are optically visible.
The distribution of such PMS stars in the Hertzprung-Russell (HR) 
diagram retains a footprint of the previous accretion phase.
The birthline is the initial loci of the PMS stars
on the HR diagram as a function of the stellar mass. 
The PMS stars move down-left toward the ZAMS line 
on the diagram thereafter.
It has been shown that the birthline for 
$\dot{M}_\ast \sim 10^{-5}~M_{\odot}/{\rm yr}$ successfully traces 
the upper envelope of observed distributions of low-mass and 
intermediate-mass PMS stars in the HR diagram 
\citep{PS92, NM00}.
In Figure \ref{fig:hr}, we show the birthlines for 
various accretion rates. 
These tracks are calculated from the interior luminosity $L_{\ast}$ 
and the stellar radius $R_{\ast}$ at the end of the accretion, i.e., 
when the protostellar mass reaches $M_{\ast}$.
Without the accretion, contribution to the total luminosity 
is now only by the interior one $L_{\ast}$ and the effective 
temperature is given by $T_{\rm *, eff} \equiv 
(L_{\ast}/4 \pi \sigma R_{\ast}^2)^{1/4}$.
With higher $\dot{M}_\ast$, protostars have larger radii and 
their birthline shifts toward the upper-right. 
This dependence on $\dot{M}_\ast$ is consistent with previous
calculations \citep{PS92, NM00}. 

If a high accretion rate of $\dot{M}_\ast > 10^{-4}~M_{\odot}/{\rm yr}$ 
is indeed achieved in massive star formation, 
stars as massive as $10~M_{\odot}$ have not yet arrived at the 
ZAMS at the end of the accretion phase.
Thus, the presence of such massive PMS stars is peculiar to the high 
$\dot{M}_\ast$ scenario. 
These PMS stars are very luminous with $L_{\ast} > 10^4 L_{\odot}$, 
and their effective temperature is much lower than that of the ZAMS stars. 
Although some candidates have been reported very close to the ZAMS line,
no such object has yet been firmly detected \citep[e.g.,][]{HHC97}. 
One explanation for the lack of detection
is the very short KH timescale $t_{\rm KH}$ of such PMS stars.
For example, in the case with $\dot{M}_\ast = 10^{-3} M_{\odot}/{\rm yr}$
(run MD3), $t_{\rm KH}$ falls below $10^4$yr for $M_{\ast} > 10 M_{\odot}$ 
(Fig.~\ref{fig:lum_enuc_tsc_1em3}, bottom). 
On the other hand, from the case with 
$\dot{M}_\ast = 10^{-5}~M_{\odot}/{\rm yr}$ (run MD5)
we see that $t_{\rm KH} > 10^6$~yr ($10^5$~yr) for low- 
(intermediate-, respectively) mass protostars 
(Fig.~\ref{fig:lum_enuc_tsc_1em5}, bottom).
In addition to the small number of massive stars, such a short 
duration in the PMS phase severely limits the possibility 
of detection.

As referred to in \S~\ref{sec:intro}, some information of 
dust-enshrouded protostars has been gleaned from the light 
re-emitted in the envelope.
On the other hand, \citet{Mn98} reported 
a unique indirect observation of an embedded massive protostar.
They observed the infrared light from the Orion BN/KL nebula,
which is scattered by dust grains far away from the exciting
protostar. Direct light from the protostar is blocked by a 
torus-like structure on the line of sight toward us, but 
likely leaks in the polar direction and is scattered by the nebula. 
The color temperature of the scattered light is evaluated as 
$T_{\rm eff} \simeq 3000-5500$K from features of the absorption lines. 
The bolometric luminosity of the protostar is 
$L_{\rm bol} > 4 \times 10^4 L_{\odot}$. 
If the scattered light is from the photosphere of the star, 
this low $T_{\rm eff}$ cannot be explained by a MS star:
$T_{\rm eff}$ of a ZAMS star with the same luminosity 
is about 35000K, much higher than the above estimation.
\citet{Mn98} concluded that a very large stellar radius
of $> 300 R_{\odot}$ is needed to explain the 
low effective temperature. 
Stimulated by this result, \citet{Nk00} studied the 
evolution of protostars with very high accretion rates 
$\sim 10^{-2} M_{\sun}/{\rm yr}$ using a one-zone model.
However, they concluded that stellar 
radius does not exceed $30~R_{\odot}$ and then
failed to reproduce the observed low effective temperature.
In contrast, using more detailed modeling, 
we have shown that the stellar radius in fact reaches 
as large as $\simeq 100M_{\sun}$ in the cases of high accretion rate .
In Figure \ref{fig:KL}, we present evolution of 
photospheric temperature and bolometric luminosity 
for three cases with $\dot{M}_\ast \geq 10^{-3}~M_{\odot}/{\rm yr}$. 
This figure shows that the required condition is 
satisfied, for example, with the case 
of $\dot{M}_\ast \geq 6 \times 10^{-3}~M_{\odot}/{\rm yr}$ for 
$M_{\ast} \simeq 20 - 25~M_{\odot}$. 
As a conclusion, the observation can be explained  
if the accretion rate is higher than $4 \times 10^{-3}~M_{\odot}/{\rm yr}$.
The allowed mass range corresponds to the duration of the swelling of 
the protostar by the luminosity wave (see \S~\ref{ssec:md_1em3}).
Recall that this swelling is caused by very inhomogeneous entropy 
distribution: the matter near the surface receives a large amount of 
entropy temporarily.
Since this is not included in Nakano et al. (2000)'s one-zone model, 
their model gave a smaller radius than ours. 
In Appendix~\ref{sec:one_zone}, we present calibration of 
parameters for a polytropic one-zone model to include this effect 
approximately.

\section{Summary and Conclusions}
\label{sec:sum}

We have studied evolution of accreting protostars 
by solving the structure of the central growing stars 
and the surrounding accreting envelopes simultaneously.
Particular attention is paid to cases with high accretion rates of 
$\dot{M}_\ast \geq 10^{-4}~M_{\odot}/{\rm yr}$, which are
envisaged in some current scenarios of massive star formation.
The protostellar evolution at a high mass accretion rate of 
$\dot{M}_\ast = 10^{-3}~M_{\odot}/{\rm yr}$ (run MD3 in Table 1) can 
be summarized as follows:

The entire evolution is divided into four characteristic
phases;
(I) adiabatic accretion ($M_{\ast} \lesssim 6~M_{\odot}$), 
(II) swelling ($6~M_{\odot} \lesssim M_{\ast} \lesssim 10~M_{\odot}$),
(III) KH contraction 
($10~M_{\odot} \lesssim M_{\ast} \lesssim 30~M_{\odot}$), 
and (IV) main-sequence accretion 
($M_{\ast} \gtrsim 30~M_{\odot}$) phases.

A main driver of transition between the evolutionary phases 
above is a decrease in opacity in the stellar interior with increasing 
temperature, and thus protostellar mass.
Due to the fast accretion, the matter accreted onto the stellar surface 
is embedded into the interior before radiatively losing  
the entropy produced at the accretion shock.
Early in the evolution, the opacity in the stellar interior, which 
is mainly by free-free absorption, is very high owing to low temperature.
As a result of the high opacity, the star keeps a large amount of 
entropy imported by the accreted matter.
This is the adiabatic accretion phase (I).
With the decrease in opacity, 
the entropy is transported outward gradually.
When the matter near the surface temporarily 
has a large amount of entropy, 
the star expands as large as $\simeq 200~R_{\odot}$.
This is the swelling phase (II).
After that, all parts of the protostar lose energy and 
the protostar contracts. 
This is the KH contraction phase (III).
With contraction, the temperature increases and the hydrogen 
burning eventually begins.
When the nuclear burning provides sufficient 
energy to balance with that lost from the star radiatively, 
the star stops the KH contraction and reaches the zero-age main-sequence 
(ZAMS) phase at $\simeq 30~M_{\sun}$ (IV). 
 
Deuterium burning plays little role in the evolution.
Even if the deuterium burning is turned off by hand, 
the result hardly changes. 
The swelling (II) is caused by outward transfer of entropy within 
the star, which is observed as a propagating luminosity wave.

In general, at the higher accretion rate, the 
protostellar radius is larger and then 
the maximum temperature is lower
at the same stellar mass, 
owing to the higher entropy within the protostar.
As a result of the lower temperature, the onset of the 
nuclear burning is postponed to the higher protostellar mass.

For very high accretion rate 
$\dot{M}_\ast > 3 \times 10^{-3}~M_{\odot}/{\rm yr}$, 
in the course of the KH contraction,
the radiation pressure onto the inner accreting envelope 
becomes so strong that the flow is retarded before hitting 
the stellar surface.
The reduced ram pressure onto the stellar surface causes
abrupt expansion of the star.
The steady-state accretion is not possible thereafter.
At this moment, the accretion might be halted, or 
might continue at a reduced rate or in a sporadic way.
In either case, the evolution at the critical accretion
rate of $M_{\rm cr} = 3 \times 10^{-3}~M_{\odot}/{\rm yr}$ gives 
the upper mass limit of PMS stars at $\simeq 60~M_\odot$.
Further growth of the star should be finally limited by the
radiation pressure onto the outer dusty envelope. 
We have found this limit with $M_{\rm cr}$ 
gives the maximum mass of MS stars at $\simeq 250~M_\odot$. 

Such a high accretion rate has also been expected and studied for 
the first star formation in the universe.
The evolutions of primordial and solar-metallicity 
protostars are very similar at the same accretion rate.
However, the lower opacity of the primordial gas results in the 
earlier transitions between the evolutionary phases above.
For example, with $\dot{M}_\ast = 10^{-3}~M_{\odot}/{\rm yr}$,
the accreting star enters the KH contraction phase at $10~M_{\sun}$ 
in the solar metallicity case, while it enters 
at $7M_{\sun}$ in the primordial case.

Distinguishing features of those massive protostars
accreting at high rates include the large stellar radius 
sometimes exceeding $100~R_{\odot}$, and 
then low color temperature $\simeq 6000$K.
A massive protostar in Orion BN/KL nebula indeed has  
such features and is a possible candidate of such objects. 

{\acknowledgements 
The authors thank Francesco Palla, Nanda Kumar, Jonathan Tan, 
Mark Krumholz, and Shu-ichiro Inutsuka for helpful comments and discussions. 
This study is supported in part by Research Fellowships of the Japan
Society for the Promotion of Science for Young Scientists (TH) and 
by the Grants-in-Aid by the Ministry of Education, Science and 
Culture of Japan (18740117, 18026008, 19047004: KO). }

\clearpage


\appendix

\section{Method of Calculations}
\label{sec:method}

\subsection{Evolutionary Calculations}

\subsubsection{Protostar}
\label{sssec:prost}

For protostellar structure, the following four stellar 
structure equations are solved:
\begin{equation}
\left( \frac{\partial r}{\partial M} \right)_t = \frac{1}{4 \pi \rho r^2},
\label{eq:con} 
\end{equation}
\begin{equation}
\left( \frac{\partial P}{\partial M}  \right)_t = - \frac{GM}{4 \pi r^4}, 
\label{eq:mom}
\end{equation}
\begin{equation}
\left( \frac{\partial L}{\partial M} \right)_t 
= \epsilon - T \left( \frac{\partial s}{\partial t}  \right)_M ,
\label{eq:ene}
\end{equation}
\begin{equation}
\left( \frac{\partial s}{\partial M} \right)_t
= \frac{G M}{4 \pi r^4} \left( \frac{\partial s}{\partial p} \right)_T
  \left( \frac{L}{L_s} - 1  \right) C ,
\label{eq:heat}
\end{equation}
with a spatial variable of Lagrangian mass coordinate $M$.
In the above, $\epsilon$ is the energy production rate 
by nuclear fusion, $s$ is the entropy per unit mass,  
$L_s$ is the radiative luminosity with adiabatic temperature
gradient, and other quantities have ordinary meanings.
The coefficient, $C$ in equation (\ref{eq:heat}) is unity if 
$L < L_s$ (i.e., in radiative layers), and given by the 
mixing-length theory if $L > L_s$ (i.e., in convective layers). 
The momentum equation (\ref{eq:mom}) simply states the
hydrostatic equilibrium.
This is valid because the stellar mechanical timescale is generally 
much shorter than the accretion timescale.
In energy equation (\ref{eq:ene}), on the other hand, the 
non-equilibrium term $-T (\partial s/\partial t)_M$ should be included.
This is because the KH timescale can be longer than or
comparable to the accretion timescale in our calculations.
Thus, the thermal equilibrium is not generally satisfied in 
accreting protostars.
Nuclear reactions included in $\epsilon$ are 
deuterium burning and hydrogen burning via the pp-chain and CN-cycle.
The elements are homogeneously distributed
in convective layers by the convective mixing.

There are two alternative schemes to integrate equations 
(\ref{eq:ene}) and (\ref{eq:heat}), which are the explicit 
and implicit schemes:

\paragraph{Explicit Scheme}
In the explicit scheme, equation (\ref{eq:ene})
is considered as a time development equation of the entropy.
In order to update the entropy at each spatial grid
from a time step $t - \Delta t$ to $t$, 
we use $(\partial L/\partial M)_t$
evaluated at the previous time step $t - \Delta t$ and 
calculate $(\partial s/\partial t)_M$ by equation (\ref{eq:ene}).
Equation (\ref{eq:heat}) is transformed as,
\begin{equation}
L = L_s \left[
         1 + \frac{4 \pi r^4}{G M} 
             \left( \frac{\partial P}{\partial s}  \right)_T
             \left( \frac{\partial s}{\partial M} \right)_t
        \right] ,
\label{eq:lpfile}
\end{equation}
with $C=1$ (radiative).
We substitute the updated entropy gradient $(\partial s/\partial M)_t$
to equation (\ref{eq:lpfile}) and calculate the luminosity profile 
$L(M,t)$.
This scheme enables the precise integration where the heat transport among
mass cells is negligible, i.e., $(\partial L/\partial M)_t$
is much smaller than other terms in equation (\ref{eq:ene}).
For example, this quasi-adiabatic situation is created in the 
very opaque interior of the protostar.

\paragraph{Implicit Scheme}
In the implicit scheme, on the other hand, time-derivative 
term $- T (\partial s/\partial t)_M$ in equation (\ref{eq:ene}) 
is regarded as a source term to calculate the luminosity profile
by spatial integration. We simultaneously integrate equations 
(\ref{eq:ene}) and (\ref{eq:heat}),
evaluating $(\partial s/\partial t)_M$ at every increment of spatial grids.
This scheme works effectively where the heat transport
among mass cells works to redistribute entropy.
For example, this is the case near the stellar surface. 
Devote special attention to discretizing the time-derivative 
$( \partial s / \partial t)_M$ in equation (\ref{eq:ene}).
For example, consider the gas element $\dot{M}_\ast \Delta t$ 
measured from the surface in a stellar model.
This element accretes to the star from a time step 
$t - \Delta t$ to $t$, and is still in the accreting envelope 
at the previous time step $t - \Delta t$.
The accreting matter plunges to the star through the accretion shock
front, which is treated as a mathematical discontinuity in
our formulation. Describing the derivative across the discontinuity
is a problem. 
We avoid this difficulty adopting a coordinate conversion (SST80b),
\begin{equation}
 \left( \frac{\partial s}{\partial t}  \right)_M = 
  \left( \frac{\partial s}{\partial t} \right)_m 
+ \dot{M}_\ast  \left( \frac{\partial s}{\partial m}  \right)_t ,
\label{eq:relm}
\end{equation}
where $m$ is the inverse mass coordinate $m \equiv M_{\ast} - M$.
The term $(\partial s/\partial t)_m$ can be obtained only with
the entropy distribution within the shock front.
This treatment is fairly valid, because the structure
of the settling layer hardly changes over
a few time steps, observing from the shock front.

In our calculations, we properly adopt different schemes
in different evolutionary stages.
In the early phase of the evolution, we use both the explicit
and implicit schemes to construct one stellar model (e.g., SST80b); 
the explicit scheme is used in the deep interior
and is switched to the implicit scheme near the stellar surface.
First, we solve the interior with the explicit scheme;
we integrate equations (\ref{eq:con}) and (\ref{eq:mom}) outward 
with a guess of central pressure $P_c$. 
When the explicitly solved region is radiative, 
we calculate $L(M,t)$ with equation (\ref{eq:ene}). 
Once active deuterium burning begins, however, 
the explicit update of entropy sometimes makes negative 
entropy gradients $(\partial s/\partial M)_t < 0$.
Such a layer should be regarded as a convective layer,
and equation (\ref{eq:lpfile}) is not valid any more.
We temporarily change the integration method in such a layer.
We correct the entropy distribution as $(\partial s/\partial M)_t = 0$
there after the explicit update, following the 
``equal-area rule'' developed by SST80b.
Once the entropy profile is corrected, we calculate 
$(\partial s/\partial t)_M$ and obtain the luminosity distribution 
by integrating equation (\ref{eq:ene}).
As the integration approaches the surface, the term 
$(\partial L/\partial M)_t$ in equation (\ref{eq:ene}) gradually grows. 
We switch to the implicit method where $(\partial L/\partial M)_t$ 
attains $0.5 \dot{M}_\ast T (\partial s/\partial m)_t$.
We integrate all equations (\ref{eq:con}) -- (\ref{eq:heat}) with 
a guess of the entropy at the switching point $s_{\rm sw}$.
When the integration reaches the stellar surface, we check
the differences from the required boundary conditions
(see \S~\ref{sssec:accflow} below).
We improve the guessed values of $P_c$ and $s_{\rm sw}$, and 
iterate this procedure until the model satisfies the outer
boundary conditions.

The above method successfully works in the early phase of evolution.
However, only in the later stage of evolution, where the entropy 
redistribution among mass cells works even in the deep stellar interior,
we need to use another method with the implicit scheme.
This occurs when a widespread convective layer appears during 
active deuterium burning, or when radiative heat transport becomes 
efficient due to the opacity decrease in the deep interior
(see \S~\ref{sec:result} for detail).
In the fully implicit method, we integrate the equations
(\ref{eq:con}) -- (\ref{eq:heat}) outward from the center with 
a guess of the central pressure $P_c$ and entropy $s_c$. 
If the integration reaches the stellar surface, we advance to 
the same fitting process at the surface as in the above method.
If the integration diverges and fails before reaching the
surface, we adopt another shooting method to a halfway fitting point.
We additionally guess the radius $R_{\ast}$ and luminosity $L_{\ast}$ 
at the stellar surface, and integrate
backward to the fitting point. 
We adjust boundary values to match the physical quantities at 
the fitting point, which are obtained by the forward and backward 
integrations.


\subsubsection{Accreting Gas Envelope and Accretion Shock Front}
\label{sssec:accflow}

The outer boundary condition of the protostar is given 
by constructing the structure of an accreting gas envelope.
The boundary layer connecting the accreting envelope and
protostar actually has detailed structure.
The acreted gas initially hits a standing shock front and
is heated up by compression there. 
The shocked gas next enters the relaxation layer behind 
the shock front and is cooled down by radiative loss.
In our formulation, this detailed structure is not solved,
but treated as a mathematical discontinuity (SST80).
We define the following two points across the discontinuity:
a post-shock point, where gas and radiation temperature
first become equal behind the relaxation layer, and a post-shock
point ahead of the shock front.
In our calculations, the former is the outer boundary 
of the protostar, and the latter is the inner boundary of the
accreting envelope.
Physical states at each point are related by a radiative 
shock jump condition. 

Here, we focus on cases where the post-shock quantities
are given in advance by the trial outward 
integration of the protostar (see \S~\ref{sssec:prost}). 
When the outward integration fails and the shooting to a halfway
fitting point is needed, we only slightly modify the same
procedure (e.g., see PS91). 
First, we judge if the accreting envelope is optically thin or thick.
Temporarily assuming that the the accretion flow is optically 
thin and in the free-fall, velocity and mass density at 
the pre-shock point are,
\begin{equation}
u_{\rm pre} = - \sqrt{\frac{2 G M_{\ast}}{R_{\ast}} } ,
\label{eq:u_pre}
\end{equation}
\begin{equation}
\rho_{\rm pre} = \frac{\dot{M}_\ast}{4 \pi R_{\ast}^2 u_{\rm pre}} .
\label{eq:rho_pre}
\end{equation}
Ahead of the shock front, outward radiation comes from both
the stellar interior and relaxation layer behind the shock front.
Therefore, radiation temperature at the pre-shock point is written
as,
\begin{equation}
T_{\rm rad, pre} = 
\left( 
\frac{L_{\ast} + L_{\rm acc}}{4 \pi R_{\ast}^2 \sigma} 
\right)^{1/4} , 
\label{eq:t_pre}
\end{equation}
where $\sigma$ is the Stefan-Boltzmann constant, and $L_{\rm acc}$ is the
accretion luminosity defined by equation (\ref{eq:lacc})
  Using equations (\ref{eq:u_pre}) -- (\ref{eq:t_pre}), optical thickness
of the accreting envelope is estimated as, 
\begin{equation}
\tau_{\rm pre} = R_{\ast} \rho_{\rm pre} \kappa(\rho_{\rm pre}, T_{\rm pre}),
\label{eq:tau_pre}
\end{equation}
where $\kappa$ is the Rosseland mean opacity.
If $\tau_{\rm pre} < 1$, the accreting envelope is optically thin
and initial adoption of the thin envelope is verified.
The jump condition at the accretion shock front is provided by
considering the flow structure across the shock front (SST80b). 
In the optically-thin case, the post-shock temperature is related
to the stellar radius and luminosity (PS91),
\begin{equation}
T_{\rm post} = \left[ \frac{1}{4 \pi R_{\ast}^2 \sigma}   
                \left( \frac12 L_{\ast} + \frac34 L_{\rm acc} \right)
               \right]^{1/4} . 
\label{eq:tpt_thin}
\end{equation}
This is the outer boundary condition of the protostar.
The protostellar models are constructed for
the post-shock quantities to satisfy equation (\ref{eq:tpt_thin}).

If $\tau_{\rm pre} > 1$, on the other hand, the accreting envelope
should actually be optically thick. In this case, we solve 
structure of the flow inside the photosphere, i.e., radiative precursor.
First, we determine the locus of the photosphere. 
Assuming the free-fall flow outside the photosphere, 
optical depth to the photosphere $\tau_{\rm ph}$ 
can be written following equations 
(\ref{eq:u_pre}) -- (\ref{eq:tau_pre}) by substituting physical
quantities at the photosphere. The only unknown quantity is the 
luminosity at the photosphere $L_{\rm ph}$,
which is needed to calculate temperature in equations (\ref{eq:t_pre}).
We calculate this photospheric luminosity using the conservation
law of the net energy outflow rate $L_e$ through the precursor,
\begin{equation}
L_e = L_{\rm ph} - \dot{M}_\ast
      \left( h_{\rm ph} + \frac12 u_{\rm ph}^2 - 
      \frac{G M_{\ast}}{R_{\rm ph}}  \right) ,
\label{eq:le1}
\end{equation}
where $h_{\rm ph} = h(\rho_{\rm ph}, T_{\rm ph})$ is the enthalpy 
of both gas and radiation at the photosphere.
Using the known post-shock quantities, we can give the numerical value of
 $L_e$ at the bottom of the accretion flow, 
\begin{equation}
L_e = L_{\ast} - \dot{M}_\ast
      \left( 
      h_{\rm post} + \frac12 u_{\rm pre}^2 - \frac{G M_{\ast}}{R_{\ast}}  
      \right) .
\label{eq:le2}
\end{equation}
Assigning this value to $L_e$ in equation (\ref{eq:le1}),
the photospheric luminosity is written with
the infall velocity and thermal state at $r = R_{\rm ph}$. 
Consequently, $T_{\rm ph}$ and $\tau_{\rm ph}$ are given as 
implicit functions of $R_{\rm ph}$.
Therefore, we can fix $R_{\rm ph}$ in an iterative fashion 
by requiring $\tau_{\rm ph} = 1$.
Once the photospheric radius is fixed, we solve the structure of the
precursor by integrating the following equations inward from 
the photosphere, 
\begin{equation}
\rho = - \frac{\dot{M}_\ast}{4 \pi r^2 u} ,
\label{eq:cont_prec}
\end{equation}
\begin{equation}
u \frac{\partial u}{\partial r} = - \frac{G M_{\ast}}{r^2} 
                        - \frac{1}{\rho} \frac{\partial P}{\partial r}
                        + \frac{\kappa}{c} F ,
\end{equation}
\begin{equation}
\frac{\partial T}{\partial r} = - \frac{3 \rho \kappa}{16 \sigma T^3} F , 
\end{equation}
where $F$ is the radiative flux.
Using equations (\ref{eq:le1}) and (\ref{eq:cont_prec}), $F$
is written as a function of $\rho$, $u$, and $T$,
\begin{equation}
F = - \rho u \left( 
             \frac{L_e}{\dot{M}_\ast} + h + \frac12 u^2 - \frac{G M_{\ast}}{r}
             \right) .
\end{equation}
When the integration reaches the pre-shock point, a jump condition 
bridges the the pre-shock and post-shock points. 
An isothermal jump condition is applied for the optically-thick 
radiative shock front, 
\begin{equation}
T_{\rm post} = T_{\rm pre}.
\label{eq:tpt_thick}
\end{equation}
This is another outer boundary condition in addition to 
equation (\ref{eq:tpt_thin}).

\subsection{Initial Models}
\label{ssec:initial}

We construct initial models by solving their structure.
The method of calculation is based on that of evolutionary
calculations, but with slight modifications.
First, we assume the entropy distribution in the stellar interior 
as an increasing function with $M$, 
\begin{equation}
s(M) = s_{c,0} 
+ \beta \frac{k_{\rm B}}{m_{\rm H}} \frac{M}{M_{*,0}} ,
\label{eq:s0}
\end{equation}
where $s_{c,0}$ is specific entropy at the stellar center, 
$k_{\rm B}$ is Boltzmann constant, $m_{\rm H}$ is atomic mass 
unit, and $\beta~(>0)$ is a free parameter.
In the core interior, we integrate equations (\ref{eq:con}) 
and (\ref{eq:mom}) outward beginning with guessed $s_{c,0}$ 
and central pressure, $p_c$. 
According to equation (\ref{eq:lpfile}), the luminosity profile there 
is calculated using an entropy gradient obtained from (\ref{eq:s0}).
Whereas this procedure is valid in the adiabatic interior, 
entropy and luminosity profiles should be consistently calculated
in place of equation (\ref{eq:s0}) near the surface
(also see \S~\ref{sssec:prost}).
The switching point is defined as follows.
Using equations (\ref{eq:ene}) and (\ref{eq:relm}), the energy
equation can be written as,
\begin{equation}
\left( \frac{\partial L}{\partial M} \right)_t = 
 \dot{M}_\ast T \left( \frac{\partial s}{\partial M}  \right)_t ,
\label{eq:ene_ini}
\end{equation}
omitting the nuclear energy production and time-derivative terms.
Once the luminosity gradient $(\partial L/\partial M)_t$ becomes 
comparable to $\dot{M}_\ast T (\partial s/\partial M)_t$ calculated with 
equation (\ref{eq:s0}), we solve the full equations including
equation (\ref{eq:ene_ini}).
The guessed $s_{c,0}$ and $p_c$ are adjusted so that the core satisfies
outer boundary conditions, as explained in \S~\ref{sssec:accflow}.
We have confirmed that some variations of initial models only affect
subsequent protostellar evolution in a very early phase.
In this paper, we adopt $\beta = 1$ as the entropy slope in 
equation (\ref{eq:s0}).

\subsection{Opacity Tables}
\label{ssec:op}

For the opacity, we adopt OPAL tables \citep[e.g.,][]{IR96} 
with the composition given by \citet{GN93}
for temperature $T > 7000$~K. 
For $T < 7000$~K, we use other opacity tables based on calculations
by \citet{AF94}. 
Contribution from grains is excluded in 
the adopted composition \citep{AGS05, CHL06}.
In all runs, we adopt $X = 0.7$, $Y = 1 - X- Z$, and
relative abundances of heavy elements following \citet{GN93}
composition.

\section{Comparison with Previous Work \\
         -- Evolution with the Low Accretion Rate
         of $10^{-5}~M_{\odot}/{\rm yr}$ --}
\label{ap:prev}

In \S~\ref{ssec:md_1em5}, we have presented the calculated 
protostellar evolution at the low accretion rate of
$10^{-5}~M_{\odot}/{\rm yr}$ (run MD5). Whereas our numerical results
reproduce basic features obtained by previous work 
(e.g., SST80a, PS91), there are still some differences.
In this appendix, we examine causes of these differences by 
exploring some cases with different deuterium
abundances and initial models.

\subsection{Dependence on Deuterium Abundance}
\label{ssec:dab}
\begin{figure}[b]
  \begin{center}
\epsfig{ file=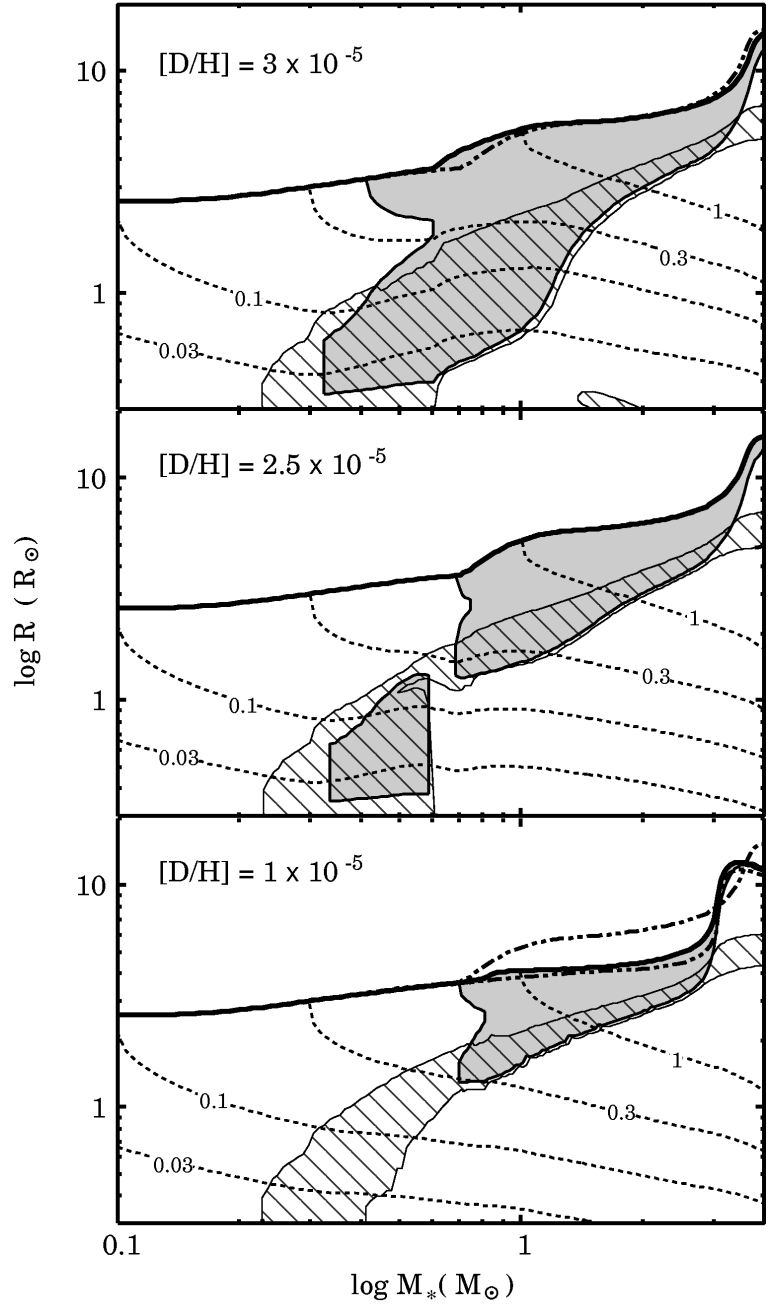,
         angle=0,
         width=4in}
\caption{The interior evolution in early phases 
for different deuterium abundances with the accretion
rate $\dot{M}_\ast = 10^{-5}~M_{\odot}/{\rm yr}$.
The deuterium abundances are [D/H] = $3 \times 10^{-5}$ 
(top; run MD5-dh3), $2.5 \times 10^{-5}$ (middle; MD5), 
and $1 \times 10^{-5}$ (bottom; MD5-dh1).
Properties of the interior structure are presented in the same
manner as in Fig.~\ref{fig:str_fdtmax_1em3}.
In each panel, the thick solid line represents the position
of the accretion shock front. The convective regions are shown 
by the gray-shaded area. 
In the top (bottom) panel, the position of the accretion shock front
for [D/H] = $2.5 \times 10^{-5}$ ([D/H]=0, run MD5-noD) is 
plotted with the dot-dashed curve for comparison. 
}
\label{fig:str_ddep}
  \end{center}
\end{figure}

The effect of different deuterium abundances 
on the stellar interior structure in early phases
is shown in Figure \ref{fig:str_ddep} for cases with 
${\rm [D/H]}=3\times 10^{-5}, 2.5\times 10^{-5}$, and $1\times 10^{-5}$.
The middle panel is the same as Figure \ref{fig:str_fdtmax_1em5} (upper), 
but reproduced for comparison.
As described in \S~\ref{ssec:md_1em5}, our run MD5
exhibits somewhat complex evolution of convective layers in the early
phase. 
A convective layer first appears at $M_{\ast} \simeq 0.3~M_{\odot}$,
just after the deuterium burning begins. 
This convective layer disappears at
$M_{\ast} \simeq 0.6~M_{\odot}$. The second convective layer emerges
just outside the former outer boundary of the first one 
at $M_{\ast} \simeq 0.7~M_{\odot}$, 
and quickly reaches the surface. 
SST80a calculated the same model, but did not observe 
such a complex evolution. In their calculation, 
the first convective layer survives and eventually incorporates 
most of the stellar interior.

This difference comes from the fact that the evolution of 
convective layers sensitively depends on the initial 
deuterium abundance. 
For example, in the case with the deuterium abundance 
[D/H] = $3 \times 10^{-5}$ 
(Figure \ref{fig:str_ddep} upper; run MD5-dh3), slightly higher than 
the fiducial value, 
the convective layer first appears at the same epoch as the fiducial case 
with [D/H] = $2.5 \times 10^{-5}$, and extends outward.
In this case, another convective layer soon emerges at the
surface and begins to extend inward.
These two convective layers finally merge at $M_{\ast} \simeq 0.6~M_{\odot}$,
and render most of the interior convective.
At $M_{\ast} \simeq 1~M_{\odot}$, the convective region contains about
97\% of the total mass. These features are very similar to
those found by SST80a, who terminated their calculation 
at $M_{\ast} = 1~M_{\odot}$.
Our calculation shows that the subsequent evolution in this case 
is almost the same as in our fiducial run MD5. 
The central radiative core gradually
expands as the stellar mass increases, and finally occupies 
most of the interior at $M_{\ast} \simeq 4~M_{\odot}$. 

This sensitive dependence on the deuterium abundance is explained
as follows. 
The expansion of convective layers is caused by 
increase of the specific entropy within it, which 
is generated by the nuclear burning at the bottom of the layer.
For continuing expansion of the convective layer, 
deuterium needs to be supplied continuously.
Fresh deuterium becomes available when the convective layer newly 
incorporates the radiative region.
This deuterium immediately spreads over the
convective layer, and is consumed for further nuclear fusion.
For sufficiently high deuterium abundance,
acquired deuterium is high enough to maintain the active nuclear burning. 
In this case, the convective layer continuously grows 
as in the case with ${\rm [D/H]}= 3\times 10^{-5}$ (run MD5-dh3). 
On the other hand, for low deuterium abundances,
the nuclear burning becomes too weak to maintain development 
of the convective layer.
The convective layer ceases to grow and disappears as in 
the case with ${\rm [D/H]}= 2.5 \times 10^{-5}$ (run MD5).
For even smaller deuterium abundance [D/H] = $1 \times 10^{-5}$ 
(Fig.~\ref{fig:str_ddep}, lower; run MD5-dh1), 
no convective layer appears even after ignition 
of the deuterium burning until at $M_{\ast} \sim 0.8~M_{\odot}$, 
when the surface convective layer finally emerges.
In this case, the subsequent mass-radius relation is 
almost same as in the case without deuterium burning (run MD5-noD)
although the surface convective layer pesists 
until $M_{\ast} \sim 3~M_{\odot}$. 
Our fiducial run MD5 is just intermediate between 
the deuterium-rich (run MD5-dh3) and -poor (MD5-dh1) cases 
presented in Figure \ref{fig:str_ddep}.
Therefore, a slight difference in input physics affects
the internal structure significantly.

\subsection{Effect of Difference in Initial Models}
\label{ssec:psini}

\begin{figure}[t]
  \begin{center}
\epsfig{ file=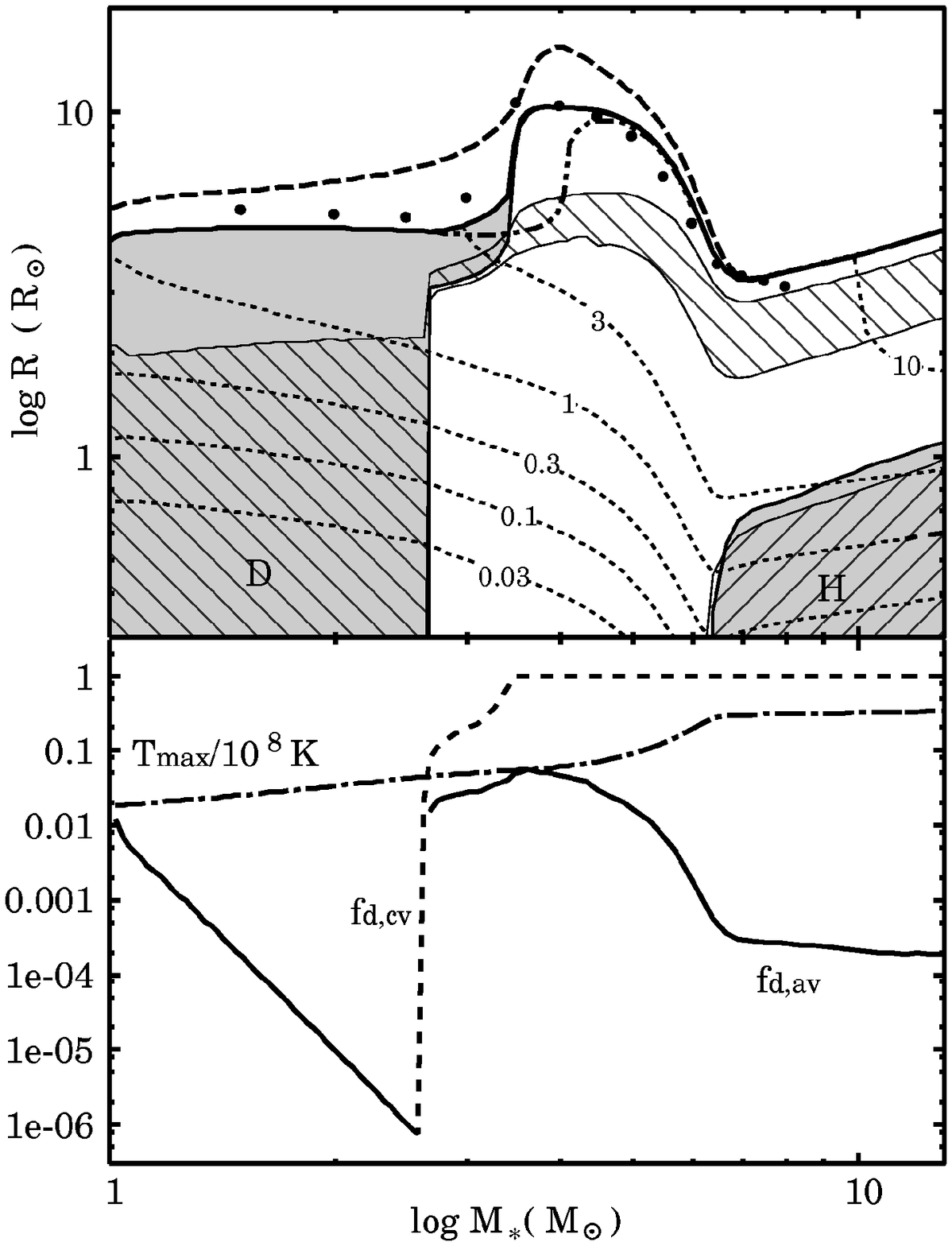,
         angle=0,
         width=4in}
\caption{
Same as Fig.~\ref{fig:str_fdtmax_1em5} but for the different
initial model calculated following PS91 
(run MD5-ps91).
In the upper panel, the filled circles denote the protostellar radius 
by PS91 with the same parameter as our fiducial case (MD5), 
which is shown by the dashed curve.
PS91 started calculation at $M_{*,0} = 1~M_{\odot}$, while the 
initial mass in our calculation is much smaller 
$M_{*,0} = 0.01~M_{\odot}$. 
The dot-dashed curve also presents the protostellar radius
in run MD5-ps91, where the deuterium 
burning is turned off at $M_{\ast} = 2.7~M_{\odot}$, 
when a radiative barrier emerges within the star (see text).
 }
\label{fig:str_fdtmax_1em5_ps}
  \end{center}
\end{figure}

Evolution of accreting intermediate-mass 
($1~M_{\odot} < M_{\ast} < 8~M_{\odot}$)
protostars have been studied in some previous work (e.g., PS91). 
Our results agree well with the previous work, 
but still show some differences even with the same accretion rate, 
which can be ascribed to different initial models. 
Whereas our initial model is a tiny radiative protostar with 
$\ll 1M_{\sun}$ (see \S~\ref{ssec:initial}), the initial model of PS91, for
example, is a $1~M_{\odot}$ fully-convective protostar. 
We confirmed that our code reproduces almost the same evolution 
as that of PS91, if we begin the calculation with the same initial model.

The initial model of PS91 is constructed in a manner similar to
that described in \S~\ref{ssec:initial}.
The homogeneous entropy distribution is adopted in place of 
equation (\ref{eq:s0}), and its value is taken from the
semi-analytic model of \citet{St88}. 
We have reconstructed the model of \citet{St88}, and obtained 
$s_{c,0} = -4.12~k_{\rm B}/m_{\rm H}$ for a $1~M_{\odot}$ star 
with our adopted opacity tables.
As in \S~\ref{ssec:initial}, we separately solve the core interior
and the surface super-adiabatic layer. 
First, we integrate the core interior outward from the center with a 
guessed value of the central pressure.  
We also guess the luminosity at the bottom of
the surface super-adiabatic layer in advance, and record the entropy 
gradient $\partial s/\partial M$ using equation (\ref{eq:heat}). 
Once the recorded $\partial s/\partial M$ attains a small finite value, 
we switch the scheme and integrate all equations 
(\ref{eq:con}) -- (\ref{eq:heat}).
The guessed quantities are adjusted for the core to satisfy the  
outer boundary conditions.
The constructed model is a fully convective star, 
whose mass and radius are $1~M_{\odot}$ and $4.2~R_{\odot}$. 

We calculate the
subsequent evolution beginning with this initial model (run MD5-ps91).
Figure \ref{fig:str_fdtmax_1em5_ps} shows the evolution of such a 
protostar. Our calculation reproduces 
the basic features of a calculation done by PS91, which adopted
the same accretion rate and accretion-shock outer boundary.
In an early phase, the star remains fully convective and 
deuterium burning occurs around the stellar center. 
Deuterium concentration significantly falls in this
phase. This is because the central temperature continuously increases,
and total burning rate $L_{\rm D}$ slightly exceeds
the steady burning rate $L_{\rm D,st}$ \citep[e.g.,][]{PS93}.
These features are somewhat different from our fiducial run MD5, 
where a radiative core persists throughout the evolution and gradually 
expands with the increase in $M_{\ast}$ by accretion 
(see Fig.~\ref{fig:str_fdtmax_1em5}). 
These differences come from different initial models.
Figure \ref{fig:str_fdtmax_1em5} also shows that mass-averaged deuterium 
concentration $f_{\rm d,av}$ does not fall much below 0.01 in our run MD5.
This is because the deuterium burning layer gradually moves to 
the outer layer, where temperature is always around $10^6$~K
and $L_{\rm D} \sim L_{\rm D,st}$. 
At $M_{\ast} = 2~M_{\odot}$, for example, $f_{\rm d,av} \sim 0.04$ in
run MD5, but only $f_{\rm d,av} \sim 10^{-5}$ in run MD5-ps91.
Therefore, our calculation predicts that intermediate-mass 
protostars have a larger amount of deuterium than assumed by PS91. 
Figure \ref{fig:str_fdtmax_1em5_ps} shows that the fully convective 
phase ends at $M_{\ast} \simeq 2.6~M_{\odot}$, when a thin
radiative layer, i.e., radiative barrier, suddenly appears 
within the star. 
The radiative barrier blocks inward convective transport of 
the accreted deuterium.
Consequently, the inner region, where deuterium has run out, 
immediately returns to being radiative.
After that, the convective layer remains only above 
the radiative core, and deuterium burning occurs 
at the bottom of the convective layer.
As in our fiducial run MD5, the protostar swells up at
$M_{\ast} \simeq 3.4~M_{\odot}$. 
PS91 have concluded that this swelling
is caused by the shell-burning of deuterium.  
To see the role of the deuterium shell-burning in the swelling, 
we also calculate evolution with cessation of the deuterium burning 
after the appearance of the radiative barrier. 
We find that the swelling occurs similarly, although it is 
slightly delayed, even without deuterium burning. 
This is caused by the outward transport of embedded entropy,
which is observed as propagation of a luminosity wave 
(see \S~\ref{ssec:md_1em3} and \ref{ssec:md_1em5}). 
Although PS91 have attributed the swelling only to the shell-burning of
deuterium, we conclude that the main driver of the swelling is always 
the propagation of the luminosity wave.
After the turn-around of the radius, the star enters the subsequent
KH contraction phase. The calculated mass-radius relation thereafter 
gradually converges with that in our fiducial run MD5.
The epoch of the Hydrogen burning, $M_{\ast} \simeq 7~M_{\odot}$, 
also agrees with that in the fuducial run.

\section{Calibration of One-zone Model}
\label{sec:one_zone}

\begin{figure}[t]
  \begin{center}
\epsfig{ file=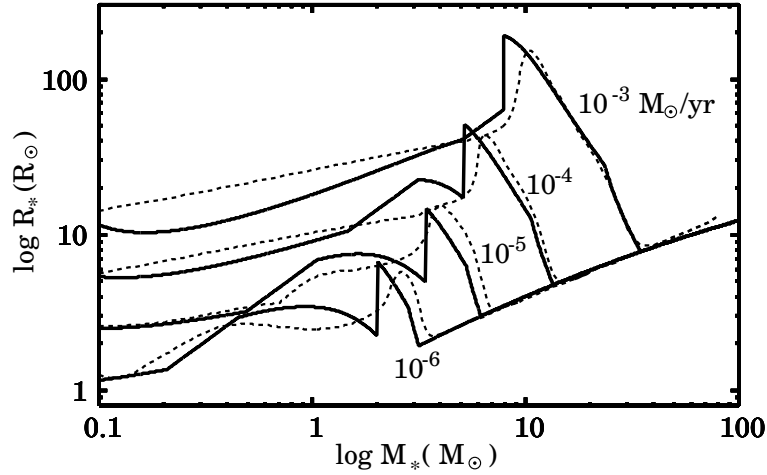,
         angle=0,
         width=4in}
\caption{Comparison between the protostellar radii calculated by 
the one-zone models based on \citet{MT03} but with parameters 
calibrated as in Sec.\ref{sec:one_zone} (solid) and by 
our numerical models (dotted). 
The cases with the accretion rates from $10^{-6}$ to 
$10^{-3}~M_{\odot}/{\rm yr}$ are shown.
 }
\label{fig:1zone}
  \end{center}
\end{figure}

While detailed numerical calculations as in this paper 
are a direct method to study the structure of accreting protostars,
one-zone modeling with the polytropic equation of state 
$P = K \rho^{1 + 1/N}$ is also a useful way. 
In the one-zone models of protostellar evolution, 
which were originally developed by Nakano, Hasegawa, \& Norman (1995), 
an energy budget of the protostar is considered.
The total energy of the star $E$ and its time-derivative $dE/dt$
are written as functions of the stellar mass $M_\ast$ and radius $R_\ast$
respectively. 
The mass-radius relation is obtained by substituting the
equation $E$ into $dE/dt$.
\citet{Nk00} applied such a model to study protostellar 
evolution at the very high accretion rate of 
$\dot{M}_\ast \sim 10^{-2}~M_{\odot}/{\rm yr}$.
\citet{MT03} improved the model by calibrating it
with numerical calculations at the accretion
rates of $\leq 10^{-4}~M_{\odot}/{\rm yr}$ (e.g., PS91).
\citet{TM04} also showed that the one-zone models can reproduce 
the mass-radius relations of primordial protostars
at the high accretion rates $\sim 10^{-3}~M_{\odot}/{\rm yr}$
\citep[e.g.,][]{OP03}.

Here, we present one-zone models reproducing our numerical 
results (Fig.~\ref{fig:1zone}).
These models are constructed following \citet{MT03} with some
improvements. 
Our numerical calculations have shown that 
protostars are initially radiative for any accretion rate.
With an accretion rate less than $10^{-4}~M_{\odot}/{\rm yr}$,
the convective layers gradually extend as deuterium burning 
becomes significant.
Since the polytropic models with index $N = 1.5$ and $N = 3$ 
approximate a fully convective and radiative star respectively
\citep[e.g.,][]{CG68}, we adopt intermediate values: 
$N = 2.5$ for $\dot{M}_\ast \le 10^{-5}~M_{\odot}/{\rm yr}$ 
and slightly larger values for higher accretion rates, 
$N = 2.5 + 0.25~\log (\dot{M}_\ast/10^{-5}~M_{\odot}/{\rm yr})$.
Initial masses and radii are arbitrarily taken as 
$M_{*,0} = 0.1~M_{\odot}$ and 
$R_{*,0} = 2.5~R_{\odot} (\dot{M}_\ast/10^{-5}~M_{\odot}/{\rm yr})^{1/3}$.
The steady deuterium burning is assumed to begin when the 
central temperature reaches $1.5 \times 10^6$~K. 
After that, we adopt $N = 1.75$, which is slightly larger 
than the fully convective value $N = 1.5$, 
to resemble our numerical result, where a radiative core 
gradually expands with the increase of the stellar mass
(see \S~\ref{ssec:md_1em5}).
Subsequently, the protostar swells up with the propagation 
of the luminosity wave. 
We include this effect into the model 
by artificially increasing the stellar radius 
by a factor of 3 to fit the calculated mass-radius relations. 
We assume that this occurs when the ratio $t_{\rm KH}/t_{\rm acc}$ 
reaches 1.75. 
Other adjustments of models are the same as those of 
\citet{MT03}. The presented one-zone models fit our 
mass-radius relations within 30\% deviation except in 
very early phases for high accretion rates.

\end{document}